\documentclass[transmag, twocolumn]{IEEEtran}
\usepackage{latexsym}
\usepackage{graphicx}
\usepackage{amsfonts,amssymb, mathtools, amsthm, mathrsfs, cuted}
\usepackage{hyperref}

\usepackage{bbm}
\usepackage{subcaption}
\usepackage{graphicx}
\usepackage{url}
\usepackage{setspace}
\usepackage{float}
\usepackage{xcolor}
\usepackage{cancel}
\usepackage{cite}

\newcommand{\bx}{\bm{X}}

\newcommand{\bh}{\bm{h}}
\newcommand{\by}{\bm{Y}}
\newcommand{\bz}{\bm{Z}}
\newcommand{\bZ}{\bm{Z}}

\newcommand{\E}{\mathsf{E}}

\newcommand{\bS}{\bm{\Sigma}}
\newcommand{\bP}{\mathbf{\Phi}}
\newcommand{\bmx}{\boldsymbol{\mathcal{X}}}
\newcommand{\mx}{{\mathcal{X}}}

\newcommand{\snr}{{\textsf{SNR }}}
\newcommand{\var}{{\textsf{Var}}}
\usepackage{lmodern,bm}                
\usepackage[T1]{sansmath} 
\usepackage[T1]{fontenc}
\usepackage[utf8]{inputenc}

\begin{document}

\title{Free-Space Optical MISO Communications With an Array of Detectors}

\author{Muhammad Salman Bashir,
\IEEEmembership{Senior Member, IEEE}, and Mohamed-Slim~Alouini, \IEEEmembership{Fellow, IEEE}
\thanks{This work is supported by Office of Sponsored Research (OSR) at King Abdullah University of Science and Technology (KAUST).}
\thanks{M.~S.~Bashir and M.~-S.~Alouini  are with the King Abdullah University of Science and Technology (KAUST), Thuwal 23955-6900, Kingdom of Saudi Arabia.  e-mail: (salman.bashir@alumni.purdue.edu, slim.alouini@kaust.edu.sa).}
}

\IEEEtitleabstractindextext{
\begin{abstract}
Multiple-input multiple-output (MIMO) and multiple-input single-output (MISO) schemes have yielded promising results in free space optical (FSO) communications by providing diversity against fading of the received signal intensity. In this paper, we have analyzed the probability of error performance of a \emph{muliple-input single-output} (MISO) free-space optical channel that employs array(s) of detectors at the receiver. In this regard, we have considered the \emph{maximal ratio combiner} (MRC) and \emph{equal gain combiner} (EGC) fusion algorithms for the array of detectors, and we have examined the performance of these algorithms subject to phase and pointing errors for strong atmospheric turbulence conditions. It is concluded that when the variance of the phase and pointing errors are below certain thresholds, signal combining with a single array of detectors yields significantly better performance than a multiple arrays receiver. In the final part of the paper, we examine the probability of error of the single detector array receiver as a function of the beam radius, and the probability of error is minimized by (numerically) optimizing the beam radius of the received signal beams.
\end{abstract}

\begin{IEEEkeywords}
Array of detectors, beam radius, equal gain combiner, maximal ratio combiner, multiple-input single-output, pointing error, strong atmospheric turbulence.
\end{IEEEkeywords}

}

\maketitle

	\section{Introduction}
	The availability of large chunks of unlicensed spectrum in the optical domain makes free-space optical (FSO) communications  an attractive solution for transmitting high data-rates for the next generation of wireless communication systems. Additionally, FSO is secure and cheaper/easier to deploy than its radio frequency (RF) counterpart. 	Typically, a laser source at the transmitter is used for signaling data, and energy detecting diodes are employed at the receiver in order to decode the signal. \emph{Intensity modulated/direct detection} (IM/DD) is a common (noncoherent) modulation scheme utilized for these systems since they do not require expensive subsystems for tracking the phase of the received signal.
	
	A receiver front end made up of an array of detectors is used typically in satellite communications in order to capture the optical signal that may be ``bouncing about'' on the array due to random fluctuations in the \emph{angle-of-arrival}. An \emph{avalanche photodiode} (APD) array is commonly used in satellites that acts as a photon counter when operated in the Geiger mode \cite{Vilnrotter:02, Vilnrotter:05, Srinivasan:16, Alerstam:18, LaserFocusWorld}. Data communications and tracking with APD detector arrays is discussed in \cite{Bashir1, Bashir2, Bashir3, Bashir7, Song:19}.

	\subsection{Motivation} \label{Motivation}
	In most research studies conducted on noncoherent \emph{multiple-input multiple-output} (MIMO) and \emph{multiple-input single-output} (MISO) FSO systems, the sufficient statistic is simply formed by first weighting the data with the channel coefficients, and then adding the resulting weighted data. For instance, as one discovers in \cite{Simon:05}, the sufficient statistic during the $i$th bit interval is $y =  x_1 h_{1} + x_2 h_{2} + n$ for the two-input single-output channel, where $x_1$ and $x_2$ are the symbols transmitted by Transmitter~1 and Transmitter~2, respectively, $h_{1}$  and $h_2$ are the coefficients of the channel corresponding to Transmitter~1 and Transmitter~2, respectively, and $n$ is a Gaussian random variable representative of thermal and shot noise. Hence, the channel  coefficients $h_{i}$ are accepted as they are, and they are simply estimated in real-time in order to decode the symbol. However, there is no study available yet that describes the effect of the beam combining from different transmitters on a single aperture, and how the phase and pointing errors can affect the the channel coefficients---and consequently---the sufficient statistic and the probability of error.
	
	By coherently combining the beam on a single receive aperture, the energy of all the beams can be concentrated on a smaller region (or a smaller number of detectors for an array of detectors), thereby leading to a more sharply focused beam with a smaller beam radius. This leads to a better signal-to-noise ratio that minimizes the probability of error. However, there is a need to study the free-space optical MISO channel in terms of the phase and pointing errors corresponding to different beams when such beams are combined on a single aperture. The limitations in this case are the phase and pointing errors which can disrupt the coherent combining process.
	
	\section{Literature Review} \label{LR}
	In this review, we have considered a number of important papers on MIMO/MISO systems in free-space optics. The authors in \cite{Letzepis:2008} have investigated the effect of multiple lasers and multiple apertures in order to mitigate scintillation for a \emph{pulse position modulation} (PPM) scheme and a Gaussian channel. In another work on MIMO FSO, the authors in \cite{Bayaki:2009} have devised the pairwise error probabilities for the single-input single-ouput (SISO) and the MIMO FSO for a Gamma-Gamma fading channel. The pairwise error probabilities are presented as a generalized infinite power series as a function of signal-to-noise ratio, and a finite version of this power series renders a fast and accurate numerical evaluation of the bit error rate in SISO/MIMO channels. Another work on MIMO FSO \cite{Tsiftsis:2009} considers the application of multiple lasers and multiple detectors in order to mitigate K-distributed atmospheric turbulence, and proposes efficient but approximate closed-form expressions for the average bit error rate of single-input multiple-output (SIMO) systems.
	
	The authors in \cite{Garcia:2009} have proposed a transmit diversity scheme for the MISO FSO based on the selection of the optical path with a greater value of scintillation. The proposed system is made up of a number of transmit lasers that project their energy on a single photodiode detector at the receiver. The proposed scheme provides a better performance than orthogonal space-time block codes and repetition codes, but requires that the channel side information be available both at the transmitter and the receiver in order to exploit selection diversity to its fullest. Another work \cite{Bayaki:2010} that deals with two transmit lasers and $N$ photodetectors furnishes expressions for pairwise error probability concerning general FSO space-time codes for a Gamma-Gamma fading channel.  Finally, the authors in \cite{Garcia-Zambrana:11} have derived the outage probability expressions for the MIMO FSO system when the FSO links suffer from strong turbulence and pointing errors. In addition to this, the authors have also optimized the beamwidth to minimize the outage probability using asymptotic expressions. 
	
The authors in \cite{Zhang:JLT:15} investigated the dual-hop relaying channel over mixed RF/FSO links which is modeled as $\eta$-$\mu / \Gamma \Gamma$ and $\kappa$-$\mu/\Gamma \Gamma$ channels. In this work, they derived the cumulative distribution function and the probability density function of the end-to-end SNR in terms of Meijer's $G$-function. They also derived the expressions of end-to-end outage probability and average bit error rate of the system. The analysis is carried out under conditions of turbulence and pointing errors. In \cite{Zhang:OJCOMS:20}, the same authors discuss the probability of error, outage probability, ergodic and effective channel capacities for a dual-hop hybrid FSO/mmWave communication system that suffers from pointing errors and atmospheric turbulence. The performance is also analyzed for different relaying techniques and fading conditions of the mmWave RF link. Finally, in \cite{Zhang:JASC:15}, the ergodic capacity expressions in the form of Meijer's $G$- and Fox's $H$-functions of a MIMO FSO system with an equal gain combining (EGC) scheme are derived.
	
	The authors in \cite{Simon:05} have proposed a modified Alamouti coding scheme for intensity modulated direct detection FSO channels. They have concluded that their proposed scheme produces a diversity of order two for a channel corresponding to two transmitters and one receiver, and the two transmitted symbols can be detected independently of each other. The extension of their proposed scheme to multiple transmitters and receiver is also considered in the same paper.

	For a discussion on  pointing error with a single detector receiver in free-space optical communications, the interested readers may refer to \cite{Farid, Mai, issaid_TWC_2017, ansari_WCOML_2013}.

	\section{Contributions, Scope and Organization of This Paper}
	\subsection{Contributions}
	As discussed in Section~\ref{Motivation}, each of the  MIMO/MISO schemes discussed in Section~\ref{LR} assume that the phase and pointing error effects are incorporated in the channel coefficients that form the sufficient statistic.  The major contribution of this paper is the analysis of probability of error of a MISO scheme when the channel suffers from phase and pointing errors. In this regard, we have derived the probability of error expressions for the MISO channel that employs a single detector array at the receiver, and its performance is compared with the scenario when multiples arrays are used at the receiver (one detector array for each beam). Both the MRC and EGC receivers are considered in the probability of error derivations. Furthermore, the beam radius is optimized in order to mitigate the effect of pointing error.
	
	Even though we have considered an array of detectors in this paper, the same analysis generalizes to any number of detectors in the array, and therefore, to a single detector  as well.
	
	\subsection{Scope}
\begin{figure}
	    \centering
	    \includegraphics[scale=0.7]{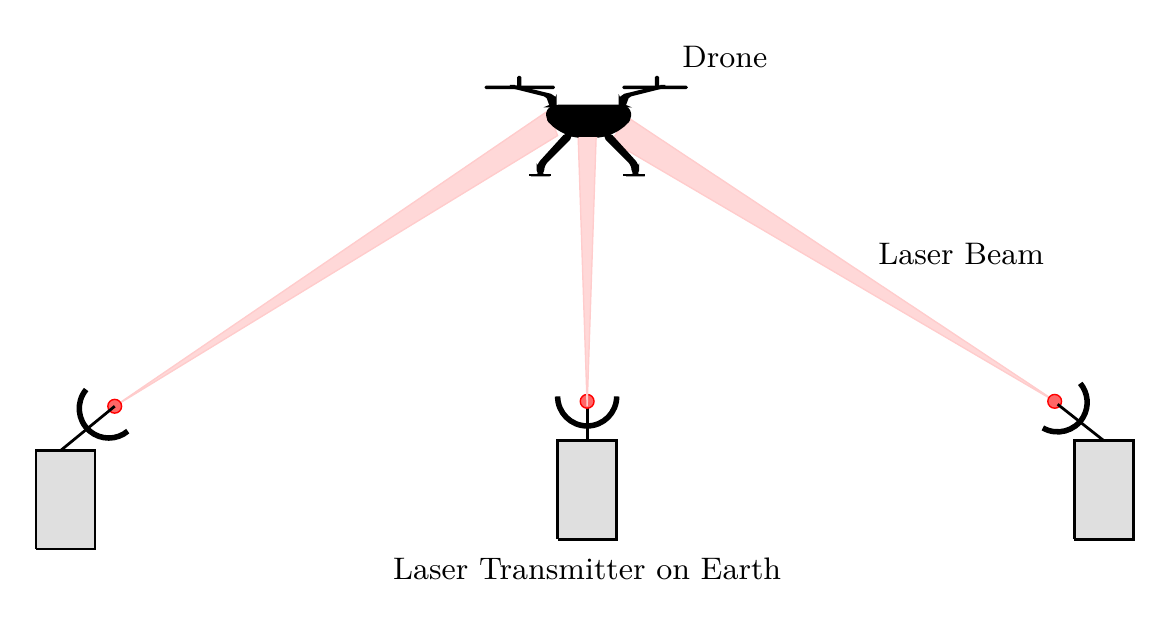}
	    \caption{In this figure, data is transmitted from ground by shooting laser to an airborne terminal (drone) from three different locations.}
	    \label{pic3}
	\end{figure}

	This article mainly studies the effect of pointing and phase errors on the performance of an FSO MISO channel. That being said, the focus of this paper is not on the space-time coding such as the Alamouti schemes that are used for a MISO channel \cite{Simon:05}. In this paper, the total transmitted laser power is split into $N$ smaller beams or channels where each beam carries the same data, and the beams are transmitted from different locations in space in order to create spatial diversity.  Fig.~\ref{pic3} depicts a drone that is being fed data from three different locations on earth. 
	\begin{figure}
	    \centering
	    \includegraphics[scale=0.5]{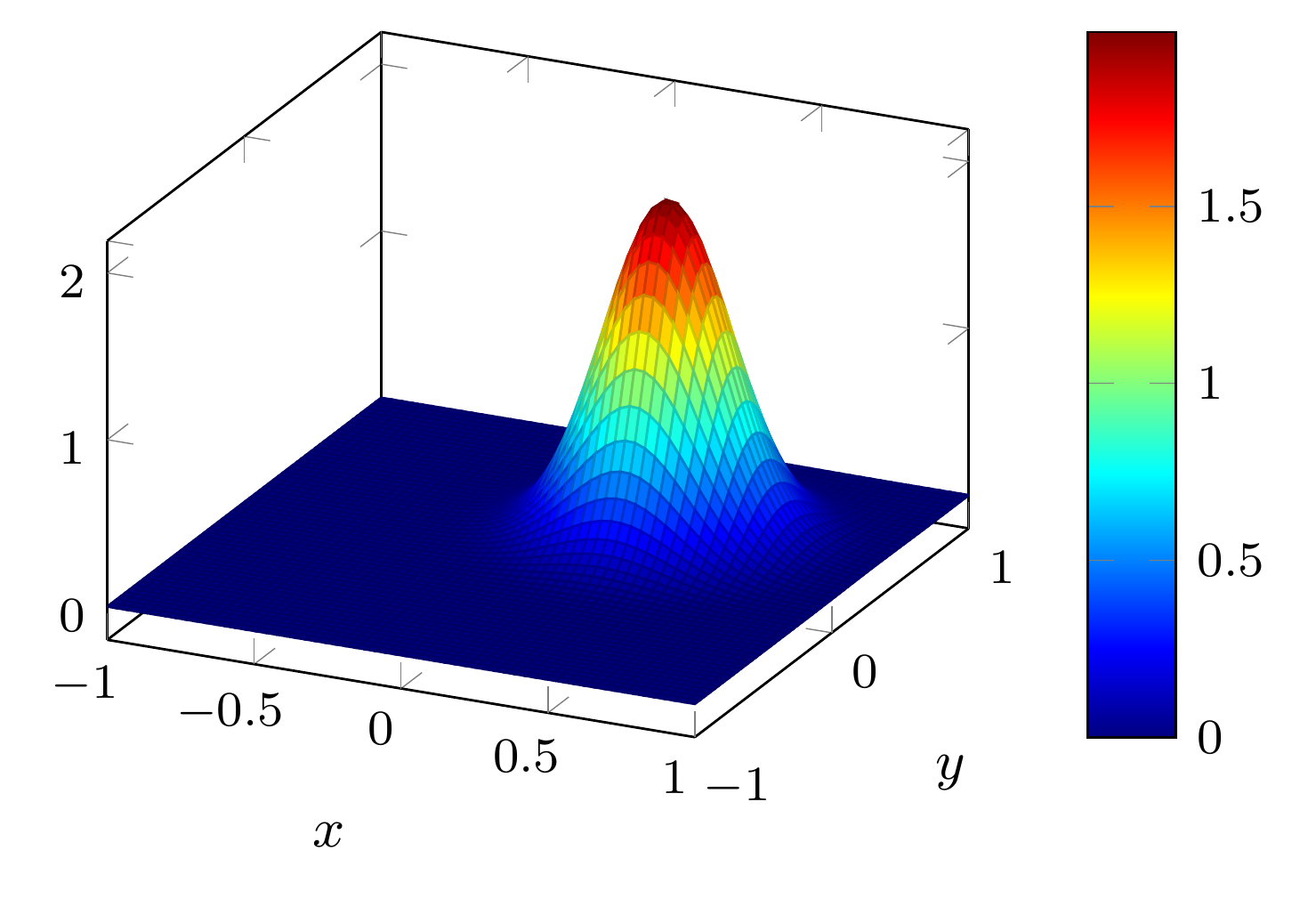}
	    \caption{The intensity of the beam on the detector array is modeled by a circularly symmetric Gaussian density function.}
	    \label{pic4}
	\end{figure}
	
	At the receiver, an array of detectors is used to detect the signal. Such an array provides a better probability of error performance as opposed to a single detector of same area as the array\footnote{The area of each detector in the array is uniform. The shape of the array is a square and the shape of each detector is also a square.} \cite{Bashir5}. This is our motivation for studying the performance of an array of detectors instead of a single detector for the MISO FSO scheme in this paper. 
	
	 The intensity of the airy pattern on the detector array is approximated by a Gaussian profile in two dimensions (see Fig.~\ref{pic4}). When a single detector array is used instead of $N$ arrays (one array for each beam), all the beams have to be combined/projected on the same array. The advantage of using a single array is a higher resulting SNR since only noise from the detectors of a single array enters the detection process, as opposed to noise from the detectors of each of the array for an $N$-arrays scenario. In other words, the SNR improves by a factor of $N$ when a single detector array is employed\footnote{This is especially true for an equal gain combiner.}. However, communications with a single detector array has its (following) drawbacks: 
	\begin{enumerate}
	    \item The different beams from the same laser travel different paths before they reach the receiver. Hence, the phase difference between different beams has to be corrected (using an interferometer) before they are combined (coherently) at the receiver. Hence, we have to deal with the \emph{phase error} of different beams in this scenario.
	    \item The airy patterns corresponding to different beams may wander about in a random fashion on the detector array due to angle-of-arrival fluctuations. These fluctuations happen either due to atmospheric turbulence, or/and due to mechanical vibrations of the transmitter/receiver assemblies. The individual airy patterns have to be tracked and aligned with the center of the detector array. Hence, the \emph{ pointing error}\footnote{This error is defined as the Euclidean distance between the center of the spot and the center of the array.} plays an important role in the performance of a single detector array receiver, and it depends on the magnitude of the angle-of-arrival fluctuations and any inertia associated with the tracking assembly. 
	\end{enumerate}

	The following is an important set of assumptions related to this study:
	\begin{enumerate}
	    \item The ratio of the receiver aperture diameter to the field coherence length is much smaller than one \cite{Vilnrotter:05}. In this case, the airy pattern on the detector array may be approximated by a Gaussian function in two-dimensional space whose peak is fluctuating according to a (random) fading coefficient $\alpha$.
	    \item The intensity fading corresponds to strong atmospheric turbulence conditions. The fading distribution in this case is the \emph{negative exponential distribution}: $f_X(x) = \eta \exp(-\eta x) \cdot \bm{1}_{[0, \infty)}(x),$ where $X$ is the received signal intensity, and $\eta$ is the mean.
	    \item We consider a high signal energy model which is typically true for a ground-based free-space optical communications channel \cite{Simon:05}. For the high energy model, the (discrete) Poisson photon counting model converges to the (continuous) Gaussian representation. Therefore,  the sum of \emph{thermal} and \emph{shot} noise in each detector is modeled by a zero-mean Gaussian random variable with variance $\sigma^2$.
	\end{enumerate}
	
	\subsection{Organization of this Paper}
	This paper is organized as follows: Section~\ref{Prelim} provides the reader with some background material regarding the problem. Section~\ref{N_arrays} contains the probability of error analysis for a receiver that comprises $N$ detector arrays. The probability of error is computed for both the maximal ratio and the equal gain combiners. Section~\ref{1-Array} discusses the probability of error argument when a single detector array is used to combine all the beams. In this regard, Section~\ref{Synchro} discusses the ideal case---the scenario in which all the received beams are phase synchronized and perfectly aligned (each beam is pointing at the center of the array with zero pointing error). In Section~\ref{Error}, we consider  the probability of error derivation when the beams incur phase and pointing errors. Since the complexity of computing the probability of error becomes significantly large when the number of beams, $N$, or the number of detectors in the array, $M$, is large, we consider an asymptotic case (large $N$ and uniformly distributed phase error) in Section~\ref{asymptotic}. This approximation simplifies the probability expression (rids the expression's dependence on $N$), and helps us understand the effect of large $N$ on the probability of error.
	
	 Since the probability of error depends on the beam radius $\rho$, we devote Section~\ref{Optimize} to the optimization of probability of error as a function of beam radius. Section~\ref{Experiments} explains the simulation results and Section~\ref{Conc} sums up the conclusions of this study.

	\section{Model Preliminaries }\label{Prelim}
	 The intensity of signal pulse falling on an array of detectors is approximately modeled by a Gaussian profile \cite{Snyder} as follows (see Fig.~\ref{pic4}):
	\begin{align}
	\lambda_s(x,y)& \triangleq \alpha \frac{\lambda_p}{\rho^2} \exp\left( \frac{-(x-x_0)^2 -(y-y_0)^2}{2 \rho^2}  \right) \cdot \boldsymbol{1}_{\mathcal{A}}(x, y), \label{intensity}
	\end{align}
	where $\lambda_p$ is peak intensity in Joules$/ \text{m}^2/\text{s}$ , $\rho$ is the beam radius in meters, $(x_0, y_0)$ is the beam center location, $\boldsymbol{1}_{{A}}$ is the \emph{indicator function} over some set $A$, and $\mathcal{A}$ is the region corresponding to the array of detectors. The quantity $\alpha$ is the intensity fading parameter that is governed by the turbulence condition of the atmosphere: $\alpha \triangleq h^* h$, where $h$ is the complex channel gain. In this case, the electric field on the detector array due to the $i$th beam amounts to 
	\begin{align}
&E_i(x, y)  = |E_i(x, y)| e^{j\phi_i} = \sqrt{\lambda_s^{(i)} (x, y)} e^{j\phi_i} \nonumber \\
&= \|h_i\| e^{j \phi_{h_i}} \sqrt{\beta_i} \frac{\sqrt{\lambda_p}}{\rho} \exp\left( \frac{-(x-x_i)^2 -(y-y_i)^2}{4 \rho^2}  \right) \nonumber \\
&\times e^{-j2\pi(u_i x + v_i y)}\cdot \boldsymbol{1}_{\mathcal{A}}(x, y), \label{intensity3}
\end{align}
	where $\lambda_s^{(i)}(x,y)$ corresponds to the intensity function of the $i$th beam,  $\|h\|$ represents the $L^2$ norm of vector $h$, $h_i, i = 0, \dotsc, N-1$, stands for the channel coefficient due to pointing errors (beam wander) and/or fading due to turbulence, and $(u_i, v_i)$ represents the center of the beam on the beam combining lens\footnote{A beam combining lens is used in order to project multiple beams on a single detector array.} surface. The factor $\phi_{h_i}$ corresponds to the phase of the $i$th channel, and $\phi_i \triangleq \phi_{h_i}+2\pi(u_ix+v_iy)$ is the total phase associated with the $i$th beam reaching the receiver. The quantity $\beta_i$ represents the fraction of the total power transmitted in the $i$th channel at the transmitter side. A suboptimal, yet effective, scheme is to choose the $\beta_i$'s as 
	\begin{align}
	\beta_i \triangleq P_t \left(\frac{\|h_i\|^2}{\sum_{j = 0}^{N-1}\|h_j\|^2} \right), i = 1, \dotsc, N, \label{19}
	\end{align}
	 where $P_t$ is the total transmitted power. From \eqref{19}, we note that a larger fraction of power is transmitted in the channels with larger channel coefficients.
	 
	 The factors $u_i$ and $v_i$ in \eqref{intensity3} correspond to the phase difference in space due to the physical separation between each of the beams being combined. Let us assume that the different beams are being combined by a single lens, and that the beam radii are much smaller than the diameter of the lens. Due to the Fourier transform property of convex lens \cite{Tyson:2014, Jutamulia:2002}, the electric field at the focal length\footnote{We assume that the array is located at a distance of one focal length from the lens.} of the lens is the (scaled) Fourier transform of the field impinging on the lens. If $(u_i', v_i')$ is the point on the lens where the $i$th beam is centered, then the electric field corresponding to the same beam on the detector array undergoes a change in phase in space. Thus, if $E(x', y') \xleftrightarrow{\mathscr{F}} \mathscr{E}(x, y)$ form Fourier transform pairs, then the electric field impinging on the lens, $E(x', y')$, and the electric field at the focal point of the lens, $\mathscr{E}(x,y)$, are related by $E(x',y')\xleftrightarrow{} \mathscr{E}\left( \frac{x}{\lambda f_0}, \frac{y}{\lambda f_0} \right)$. The quantity $\lambda$ is the wavelength of light and $f_0$ is the focal length of the lens (both quantities are expressed in meters). For any displacement $(u_i', v_i')$ of the impinging electric field from the center of the lens, the two electric fields are related by
	\begin{align}
	    & E(x'-u_i', y'-v_i') \xleftrightarrow{} \mathscr{E}\left( \frac{x}{\lambda f_0}, \frac{y}{\lambda f_0} \right) e^{-j 2\pi\left(u_i' \frac{x}{\lambda f_0} + v_i' \frac{y}{\lambda f_0} \right) } \nonumber \\
	    &=  \mathscr{E}\left( \frac{x}{\lambda f_0}, \frac{y}{\lambda f_0} \right) e^{-j 2\pi\left( \frac{u_i'}{\lambda f_0} x + \frac{v_i'}{\lambda f_0} y \right) } \nonumber \\
	    &= \mathscr{E}\left( \frac{x}{\lambda f_0}, \frac{y}{\lambda f_0} \right) e^{-j 2\pi\left( u_i x + v_i y \right) },
	\end{align}
	where $u_i \triangleq \frac{u_i'}{\lambda f_0}$ and $v_i \triangleq \frac{v_i'}{\lambda f_0}$. Thus, \eqref{intensity3} follows.

	\section{Performance Analysis of Multibeam FSO With Multiple Detector Arrays}\label{N_arrays}
	
	The output vector of the entire system ($N$ arrays) is given by
	\begin{align}
	\bm{Y} =   \bm{X} +  \bm{V} \label{1}
	\end{align}
	where $\bm{Y} \triangleq \begin{bmatrix}
	\bm{Y}_0 & \bm{Y}_1 & \dotsm & \bm{Y}_{N-1}
	\end{bmatrix}^\intercal$, $\bm{X} \triangleq \begin{bmatrix}
	\bm{X}_0 & \bm{X}_1 & \dotsm & \bm{X}_{N-1}
	\end{bmatrix}^\intercal$, and $\bm{V} \triangleq \begin{bmatrix}
	\bm{V}_0 & \bm{V}_1 & \dotsm & \bm{V}_{N-1}
	\end{bmatrix}^\intercal$. In this notation, the subscript $i$ in $\bm{Y}_i$, $\bm{X}_i$ and $\bm{V}_i$ denotes the output signal, the input signal, and the Gaussian noise vector, respectively, corresponding to the $i$th detector array. Thus, $
\bm{Y}_i =  \bm{X}_i + \bm{V}_i,$
where $\bm{Y}_i \triangleq \begin{bmatrix}
Y_i^{(0)} & Y_i^{(1)} & \dotsm & Y_i^{(M-1)}
\end{bmatrix}^\intercal$,  $\bm{X}_i \triangleq \|h_i\|^2 \beta_i\begin{bmatrix}
x_i^{(0)} & x_i^{(1)} & \dotsm & x_i^{(M-1)}
\end{bmatrix}^\intercal$ and $\bm{V}_i \triangleq \begin{bmatrix}
V_i^{(0)} & V_i^{(1)} & \dotsm & V_{i}^{(M-1)}
\end{bmatrix}^\intercal$. 

For $N$ detector array scenario, there is a single beam for each detector array, and that beam is projected at the center of the focusing lens $(u_i = v_i = 0)$. Thus, let us define the following quantity
\begin{align}
&x_i^{(m)}  \triangleq  \iint_{A_m} \left\| e^{j \phi_{h_i}}  \frac{\sqrt{\lambda_p}}{\rho} e^{ -\frac{(x-x_i)^2 -(y-y_i)^2}{4 \rho^2}  } \right\|^2  dx \, dy \nonumber \\
&=  \iint_{A_m}  \frac{{\lambda_p}}{\rho^2} e^{ -\frac{(x-x_i)^2 -(y-y_i)^2}{2 \rho^2}  }   \  dx \, dy.  \label{18}
\end{align}
 
 Additionally, we assume that the noise vectors are uncorrelated  for each detector array, and that the noise in each element of a single detector array is also uncorrelated with every other element. Thus, $\bm{V} \sim \mathcal{N}(\boldsymbol{0}, \bS)$, where the vector $\boldsymbol{0}$ is an $M N \times 1$ vector of zeros, and the matrix $\bS =  \sigma^2 \mathbf{I}_{MN}$, where $\mathbf{I}_m$ is an $m \times m$ identity matrix for some positive integer $m$.

\subsection{Maximal Ratio Combiner for $\mathcal{M}$-Ary PPM}

	In a pulse position modulation scheme, the information about the transmitted symbol is encoded in the position of the transmitted light pulse in a certain time interval (a symbol period). A PPM symbol period is divided into a number of slots, and the location of the pulse in a given slot represents the symbol. In this section, we  consider a $\mathcal{M}$-PPM scheme where $\mathcal{M} \triangleq 2^n$ for $n$  a positive integer. Let us assume that a symbol $j$ is transmitted where $0 \leq j < \mathcal{M}$. It can be shown that the maximum likelihood detector in this case shall decide in favor of some symbol $j$ if 
	 $\mathcal{L}\left( \bz^{(j)} \right) > \mathcal{L} \left( \bz^{(i)} \right)$
	 for every $i \neq j$. The quantity $\bz^{(k)}$ denotes the data vector obtained during the $k$th time slot of the $\mathcal{M}$-ary PPM scheme. The quantity $\mathcal{L}(\bm{Z})$ is the likelihood ratio which is defined as $\displaystyle \mathcal{L}(\bm{Z}) \triangleq \frac{p_1(\bm{Z})}{p_0(\bm{Z})}$ for any data vector $\bm{Z}$. The quantity $p_1(\bm{Z})$ corresponds to the conditional density function of $\bm{Z}$ given that the signal is present in a given slot of PPM, and $p_0(\bm{Z})$ indicates the conditional density function corresponding to the ``noise-only'' slot. 
	 
	 Let us assume that symbol $j$ out of the $\mathcal{M}$ PPM symbols  is transmitted to the receiver. For a maximum likelihood detector, the probability that symbol $j$ is (correctly) decided at the receiver is 
 \begin{align}
 P(c | j) = P\left(   \left\{ \ln \mathcal{L}(\bm{Y}_{s}) > \ln  \mathcal{L}(\bm{Y}_{n}) \right\} \right)^{\mathcal{M}-1}
 \end{align}
 for $i \neq j$, and $\bm{Y}_s$ corresponds to the observations when the laser is turned on (signal plus noise scenario), and $\bm{Y}_n$ corresponds to the data set when the laser is turned off (noise-only scenario).  Thus,
  \begin{align}
&\ln \mathcal{L}(\bm{Y}_s) = \ln \frac{p_1(\by_s)}{p_0(\by_s)} = \ln p_1(\by_s) - \ln p_0(\by_s)\nonumber \\
&= -\frac{1}{2}(\by_s -  \bx)^\intercal \bS^{-1} (\by_s -  \bx) + \frac{1}{2} \by_s^\intercal \bS^{-1} \by_s.
 \end{align}
Therefore, we have that
\begin{align}
&P\left(   \left\{ \ln \mathcal{L}(\bm{Y}_{s}) - \ln  \mathcal{L}(\bm{Y}_{n}) > 0 \right\} \right) \nonumber \\
&= P\left( \left\{ - (\by_s -  \bx)^\intercal \bS^{-1} (\by_s - \bx) + \by_s^\intercal \bS^{-1} \by_s \right. \right.  \nonumber \\
& \left. \left. + (\by_n -  \bx)^\intercal \bS^{-1} (\by_n -  \bx) - \by_n^\intercal \bS^{-1} \by_n  > 0 \right\} \right) \nonumber \\
& = P \left( \left\{    \by_s^\intercal  \bx - \by_n^\intercal   \bx   > 0 \right\}  \right) \nonumber \\
&= P\left( \left\{ \sum_{i=0}^{N-1} \sum_{m=0}^{M-1} \left( Y_{i,s}^{(m)} - Y_{i,n}^{(m)}\right) \| h_i\|^2 \beta_i  x_i^{(m)} > 0 \right\}  \right)
\end{align}
	where $x_i^{(m)}$ is given in \eqref{18}. The quantity $\left( Y_{i,s}^{(m)} - Y_{i,n}^{(m)}\right) \| h_i\|^2 \beta_i x_i^{(m)}$ is a Gaussian random variable with mean $\left(\| h_i\|^2 \beta_i x_i^{(m)} \right)^2$, and variance $2 \sigma^2 \left( \| h_i\|^2 \beta_ix_i^{(m)} \right)^2$. Let us denote $Z \triangleq \sum_{i=0}^{N-1} \sum_{m=0}^{M-1} \left( Y_{i,s}^{(m)} - Y_{i,n}^{(m)}\right)  \| h_i\|^2 \beta_i x_i^{(m)}$. Then, $Z$ is normal with mean, 
	$\E[Z] =    \sum_{i=0}^{N-1} \sum_{m=0}^{M-1}  \left( \| h_i\|^2 \beta_i x_i^{(m)} \right)^2$, and variance, $ \var[Z] = \sum_{i=0}^{N-1} \sum_{m=0}^{M-1} 2 \sigma^2 \left( \| h_i\|^2 \beta_i x_i^{(m)} \right)^2.$ Therefore,  
	\begin{align}
	&P\left(   \left\{ \ln \mathcal{L}(\bm{Y}_{s}) - \ln  \mathcal{L}(\bm{Y}_{n}) > 0 \right\} \right) = P\left( \{ Z > 0\} \right) \nonumber \\
	&= Q \left(-  \frac{\sum_{i=0}^{N-1} \sum_{m=0}^{M-1}  \left( \| h_i\|^2 \beta_i x_i^{(m)} \right)^2}{\sigma \sqrt{\sum_{i=0}^{N-1} \sum_{m=0}^{M-1} 2  \left( \| h_i\|^2 \beta_i x_i^{(m)} \right)^2} }   \right).
	\end{align}
	 If we assume an equiprobable symbol scheme for $\mathcal{M}$-PPM, i.e., $P(e)= 1 - P(c) = 1 - P(c|j),$ we have a final expression for the probability of error: 
	 \begin{align}
	 {P(e|\bh) \! = \! 1 \!-\! \left( \! Q \! \left(- \sqrt{  \frac{\sum_{i=0}^{N-1} \sum_{m=0}^{M-1}  \left( \| h_i\|^2 \beta_i x_i^{(m)} \right)^2}{2\sigma^2  } }   \right)      \right)^{\!\! \!\mathcal{M}-1}}, \label{MRC_Pe}
	 \end{align}
	 where $\bh$ represents a realization of channel coefficient vector : $\bh \triangleq \begin{bmatrix}
	 h_0 & h_1 & \dotsm & h_{N-1}
	 \end{bmatrix}^\intercal$. The probability of error is then 
	 \begin{align*}
	 P(e) = \int_{-\infty}^\infty\dotsm \int_{-\infty}^{\infty} P(e|\bh) f(\bh)\, d\bh,
	 \end{align*}
	where $f(\bm{h})$ is the probability density function of $\bm{h}$.
	 
	 \subsection{Equal Gain Combiner for $\mathcal{M}$-Ary PPM} \label{EGC1}
	 The probability of error for an EGC receiver is given by 
	 \begin{align}
	 P(e|\bh) = 1 - P(c|\bh) = 1 - \left(  P\left(  \left\{ Z_s - Z_n > 0 \right\}  \right)  \right)^{\mathcal{M}-1},
	 \end{align}
	 where $Z_s$ and $Z_n$ correspond to the sufficient statistic for the ``signal plus noise'' and ``noise only'' scenarios, respectively. For the EGC, $Z_s \triangleq \sum_{i=0}^{N-1} \sum_{m=0}^{M-1} Y_{i,s}^{(m)}$, and $Z_n \triangleq \sum_{i=0}^{N-1} \sum_{m=0}^{M-1} Y_{i,n}^{(m)}$. Then, the probability of error is
	 \begin{align}
	 P(e|\bh)= 1 - \left(P \left( \left\{  \sum_{i=0}^{N-1} \sum_{m=0}^{M-1} \left( Y_{i,s}^{(m)} - Y_{i,n}^{(m)} \right) > 0 \right\}   \right) \right)^{\mathcal{M}-1},
	 \end{align}
	 where $Y_{i,s}^{(m)} \sim \mathcal{N}\left( \| h_i\|^2 \beta_i x_i^{(m)}, \sigma^2  \right)$ and  $Y_{i,n}^{(m)} \sim \mathcal{N}\left(  0, \sigma^2  \right)$. After a few manipulations, we arrive at the probability of error:
	 \begin{align}
	{ P(e|\bh) = 1 - \left(  Q \left(   - \frac{\sum_{i=0}^{N-1}  \sum_{m=0}^{M-1} \| h_i\|^2 \beta_i x_i^{(m)}  }{\sqrt{2\sigma^2 NM}}     \right)    \right)^{\mathcal{M}-1}}. \label{EGC_Pe}
	 \end{align}

	\section{Multibeam FSO With a Single Detector Array: Perfect Phase Synchronization and Zero Pointing Error Case}\label{1-Array}
	
	\begin{figure*}
	\centering
	    
	    \includegraphics[scale=0.65]{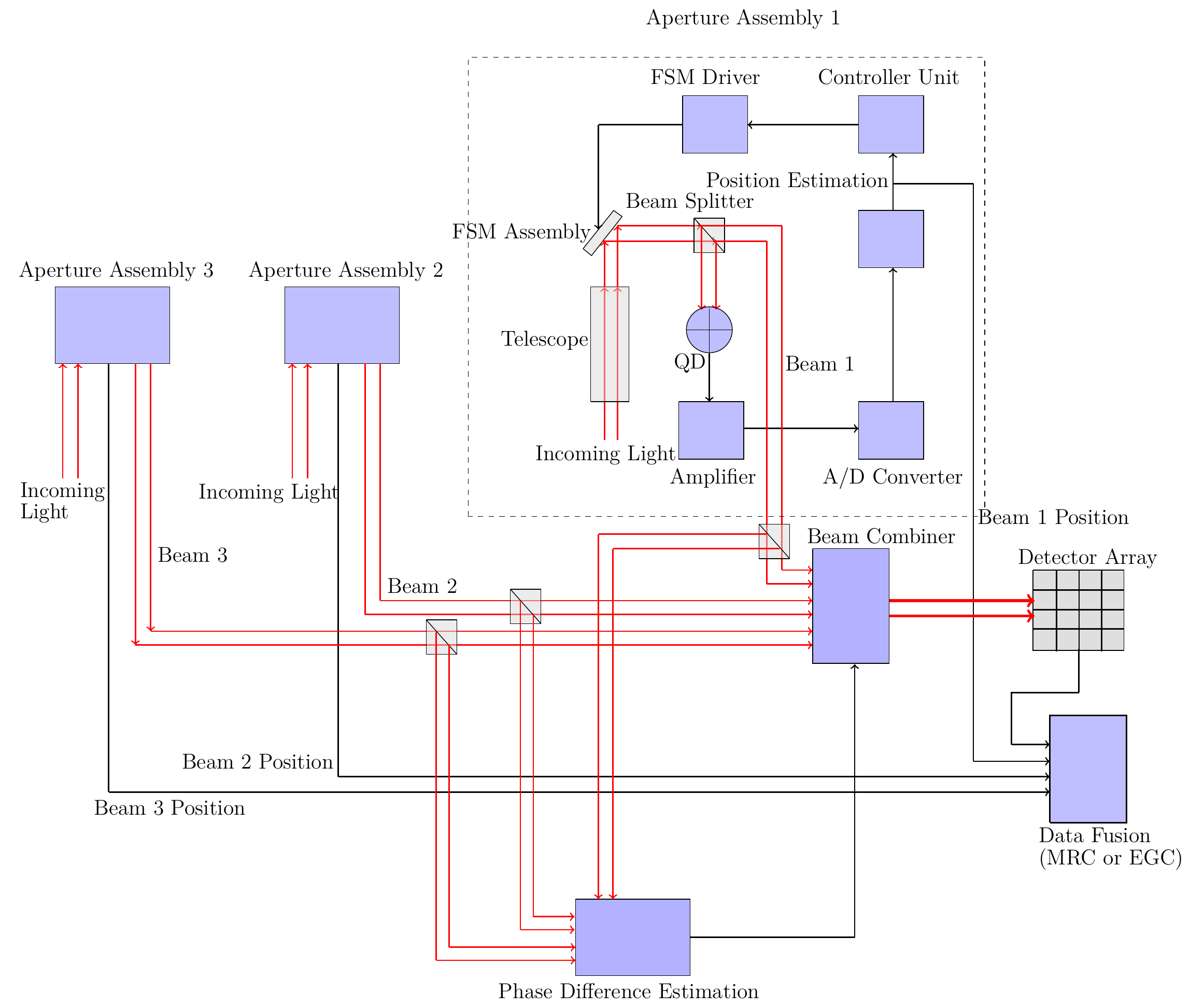}
	    \caption{This block diagram depicts beam tracking and data detection mechanism when a single detector array is employed at the receiver. The number of beams $N = 3$.}
	    \label{pic2}
	\end{figure*}
	In this case, the total power is split into $N$ different beams that are transmitted on uncorrelated channels towards a receiver that contains a single detector array. In this case, the overall intensity at the detector array is given by
	\begin{align} \label{intensity1}
	&\lambda_s(x,y) = \left( \sum_{i = 0}^{N-1} \| E_s^{(i)}(x,y)\| e^{j \phi_i} e^{-j 2\pi(u_i x + v_i y)} \cdot \bm{1}_{A_m}(x,y)\right) \\ \nonumber  &\times \left( \sum_{\ell = 0}^{N-1} \| E_s^{(\ell)}(x,y)\| e^{j\phi_\ell}  e^{-j 2\pi(u_\ell x + v_\ell y)} \cdot \bm{1}_{A_m}(x,y)\right)^* 
	\end{align}
	The output of the $m$th detector of the array is 
	$ Y^{(m)} =  X^{(m)} + V^{(m)}$,
	where
	$X^{(m)} \triangleq  \iint_{A_m} \lambda_s(x,y)\, dx \, dy$ and $V^{(m)} \sim \mathcal{N}(0, \sigma^2)$.

	\subsection{Pointing Error and Beam Alignment on a Single Array}
	
	\begin{figure}
	    \centering
	    \includegraphics[scale=0.7]{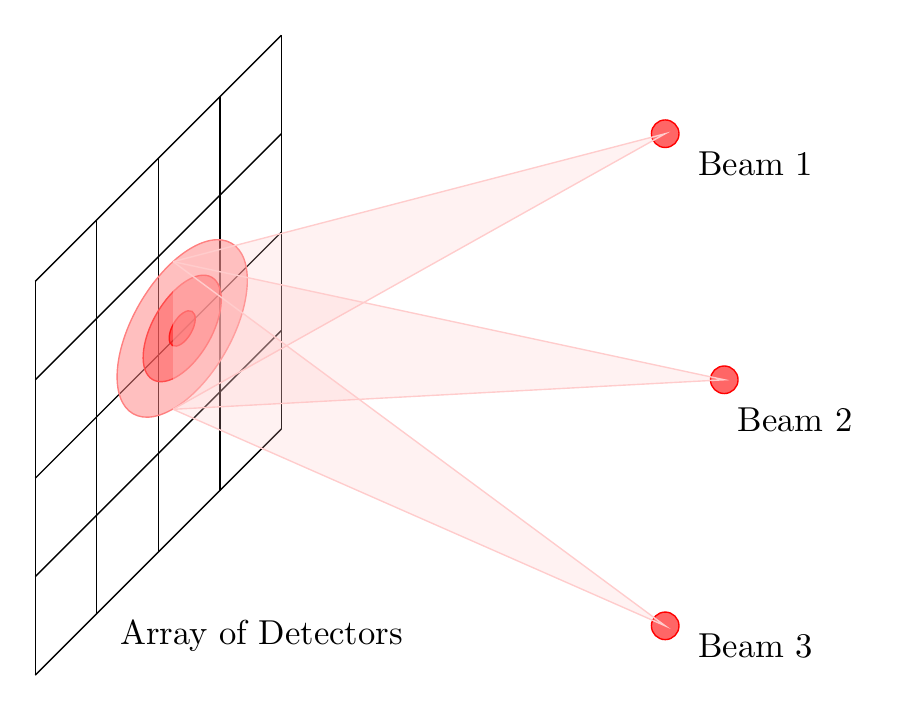}
	    \caption{This figure depicts three laser beams impinging and overlapping on an array of detectors at the receiver.}
	    \label{pic1}
	\end{figure}
	If the beams do not interfere---that is, they do not overlap with each other---then the probability of error performance of the equal gain combiner will be the same as the performance obtained in \eqref{5}. However, interference or overlap of the beams results in fluctuation of the resulting  signal intensity due to phase error which in turn affects the probability of error. Nevertheless, as we will see in Section~\ref{Experiments}, if the phase error between different beams is minimized, then the probability of error improves significantly.
	
	Assuming that there is a feedback loop in place that drives the fast steering mirror (FSM) assembly in order to keep the beams aligned with the center of the array, the deviation of any beam from the center of the array is modeled as a Gauss-Markov Process \cite{Bashir2}. This deviation is termed as the \emph{pointing error}. Let us denote the pointing error at discrete-time $n$ by the two dimensional random vector $\bmx_n^{(i)}$, and its realization by $\mx_n^{(i)} \triangleq \begin{bmatrix}
	x_n^{(i)} & y_n^{(i)}
	\end{bmatrix}$. Then, the Gauss-Markov model of the beam position evolution is \cite{Bashir2}
	\begin{align}
	\bmx_n^{(i)} = \mathbf{\Phi}\bmx_{n-1}^{(i)} + \bm{U}_n^{(i)} + \bm{W}_n^{(i)}, \label{GM}
	\end{align}
	where $\mathbf{\Phi}$ is a $2\times 2$ state transition matrix and  $\bm{W}_n^{(i)}$ refers to a $2 \times 1$ zero-mean Gaussian noise vector. The positive integer $n$ corresponds to the time index, and $i$ stands for the beam index. We assume that  $\bm{W}_k^{(i)} \perp \bm{W}_\ell^{(i)}$ for $k \neq \ell$, and $\bm{W}_n^{(i)} \perp \bm{W}_n^{(j)}$ for $i \neq j$. Additionally, for the sake of simplicity, we assume that the statistics of $\bm{W}_n^{(i)}$ is the same as $\bm{W}_k^{(j)}$ for all $n, k, i$ and $j$. Thus $\bm{W}_n^{(i)}$ is a zero-mean Gaussian vector with covariance matrix $\mathbf{\Sigma}_W \triangleq \sigma_W^2 \mathbf{I}_2$ for each $i$ and $n$. The quantity $\bm{U}_n^{(i)}$ represents the control input at time $n$ for the $i$th beam.
	
	The eigenvalues of $\mathbf{\Phi}$ depend on the physical parameters of the channel that leads to a (random) beam wander. Let us assume that the control input $\bm{U}_n^{(i)}$ is applied after every $n_0$ time units with the help of a FSM assembly. The job of the control input is to align the beam center to the center of the array. If $n_0$ is too large (delay is large between two successive control inputs), then there is a chance that a significant portion of the beam energy will leave the array which results in an outage at the receiver. If $n_0$ is too small, then we are activating the tracking and FSM assembly too frequently which results in a waste of energy at the receiver. Hence, $n_0$ has to be chosen carefully. After fixing $n_0$, the control input $\bm{U}_n^{(i)}$ is given by
	\begin{align}
	    \bm{U}_n^{(i)}& \triangleq \begin{dcases} -{\bm{\Phi} \hat{\bm{\mathcal{X}}}_{n-1}^{(i)}}, \quad n= n_0, 2n_0, \dots\\
	    \bm{0}, \quad \text{otherwise},
	    \end{dcases}
	    \end{align}
	    where $\hat{\bm{\mathcal{X}}}_{n-1}^{(i)}$ is the estimate of $\bm{\mathcal{X}}_{n-1}^{(i)}$, and $\bm{0}$ is the $2\times 1$ zero vector. We assume that FSM assembly aligns the beam center to the center of the array almost instantaneously. Thus, \eqref{GM} at time instant $n_0, 2n_0, \dotsc,$ is rewritten as
	\begin{align}
	    	\bmx_n^{(i)} = \mathbf{\Phi}\left(\bmx_{n-1}^{(i)} - \hat{\bm{\mathcal{X}}}_{n-1}^{(i)}\right) + \bm{W}_n^{(i)} \text{ for } n = kn_0, \label{GM1}
	\end{align}
	where $k$ is a positive integer.
	Thus, if $\hat{\bm{\mathcal{X}}}_{n-1}^{(i)} = \bm{\mathcal{X}}_{n-1}^{(i)}$, then the FSM assembly aligns the beam to the center of the array perfectly every $n_0$ time instants\footnote{Please note that for perfect alignment, we also need prior knowledge of state transition matrix $\bm{\Phi}$.}. In other words, the system resets the beam to the center of the array  every $n_0$ time units, and we can assume that $\bmx_n^{(i)}$ has the same statistics (mean and covariance matrix) for $n = n_j$ and $n = kn_0 + n_j$ where $n_j  \in \mathbb{Z}^+$ and $n_j < n_0$. As an example, $\bmx_0^{(i)}$ has the same statistics as $\bmx_{n_0}^{(i)}$, and $\bmx_{1}^{(i)}$
	has the same statistics as $\bmx_{n_0+1}^{(i)}$.
	
	Let us assume that we have perfect alignment every $n_0$ time unites. Then for $ 0 \leq n  < n_0$,  \eqref{GM} can be rewritten as 
	\begin{align}
	\bmx_n^{(i)} = \mathbf{\Phi}^n \bmx_0^{(i)} + \sum_{{\ell=1} }^n  \mathbf{\Phi}^{n-\ell} \bm{W}_\ell^{(i)}. \label{evo}
	\end{align}
	Then, if we assume that $\E[\bmx^{(i)}_0] = \boldsymbol{0}$, we have that $\E[\bmx^{(i)}_n]= \boldsymbol{0}$. Additionally, we assume that the covariance matrix of $\bmx^{(i)}_0$ is $\mathbf{\Sigma}_0 \triangleq \sigma_0^2 \mathbf{I}_2$. Therefore, the covariance matrix of $\bmx^{(i)}_n$ is
	\begin{align}
	&\bS_n  = \sigma_0^2 \bP^n  \left(\bP^n\right)^\intercal + \sum_{{k=1} }^n \sum_{{\ell=1} }^n  \mathbf{\Phi}^{n-k} \bm{W}_k^{(i)}     \left(\bm{W}_\ell^{(i)} \right)^\intercal \left( \mathbf{\Phi}^{n-\ell} \right)^\intercal\nonumber \\
	&= \sigma_0^2 \bP^n  \left(\bP^n\right)^\intercal + \sigma_W^2 \sum_{{\ell=1} }^n  \mathbf{\Phi}^{n-\ell}  \left( \mathbf{\Phi}^{n-\ell} \right)^\intercal.
	\end{align}
	In case $\bP$ is symmetric and the magnitude of its eigenvalues is less than 1, we have  a simpler form:
	\begin{align}
	\bS_n &= \sigma_0^2 \bP^{2n}     \sigma_W^2 \sum_{{\ell=1} }^n \mathbf{\Phi}^{2(n-\ell)} \overset{\text{large }n}{\approx}  \sigma_W^2 \sum_{{\ell=1} }^n \mathbf{\Phi}^{2(n-\ell)}\nonumber \\
	&= \sigma_W^2 \sum_{j=1}^{n} \bP^{2j}. \label{6}
	\end{align}

	Additionally, we note that the variance of $\bmx^{(i)}_n$ depends on two factors: the variance of the noise disturbance, $\sigma_W^2$, and the magnitude of the eigenvalues of state transition matrix $\bP$. 
	
	The magnitude of $\sigma_W^2$ depends on the sampling rate at which we track $\bm{\mathcal{X}}_n^{(i)}$. Let us assume that $\bm{\mathcal{X}}_n$ is the sampled version of a (continuous-time) \emph{Wiener process} $\bm{\mathcal{X}}(t)$. Then the increments $\bm{W}(t) = \bm{\mathcal{X}}(t+\Delta t) - \bm{\mathcal{X}}(t)$, and $\bm{W}_n$ and $\bm{W}_m$ are independent zero-mean Gaussian random variables for discrete-time indices $n \neq m$ whose variance is equal to $\Delta t$ \cite{Durrett:19}. Thus, if the beam position on the array is estimated frequently enough (for example, every symbol period), the variance $\sigma_W^2$ can be minimized significantly. The downside, however, is the large complexity involved in estimating the beam position at a high rate.  
	
	We define the \emph{coherence region} as a circle of radius $\rho$ on the detector array whose center lies on the point $(0,0)$ (this point is also the center of the array). The goal of the tracking assembly is to achieve beam alignment, i.e., the centers of different beams should be aligned with point $(0,0)$ on the array. The coherence region is denoted by $\mathcal{A}_c$. 
	
	We say that all the beams are ``aligned'' if their beam centers lie within $\mathcal{A}_c$, i.e., $(x_i, y_i) \in \mathcal{A}_c$ for $i =0, 1, \dotsc, N-1$. This is an approximate  argument since, in this case, we are assuming that the resulting interference is the same regardless of where the beam centers are located inside $\mathcal{A}_c$. Furthermore, as soon as a particular beam center leaves $\mathcal{A}_c$, we assume that its contribution to the total interference due to the remaining beams in $\mathcal{A}_c$ is negligible. 
	
	Finally, we want to distinguish between the three scenarios: i) When the beams are aligned, they overlap but there may still nonzero pointing error associated with each beam. However, the tracking assembly aligns each beam every $n_0$ time units. ii) The perfect beam alignment scenario implies that the pointing error is zero for each beam and all beam centers coincide with the center of the array. iii) When the beams are not aligned, that means that the different beams do not overlap with each other at any time.

	\subsection{Beam Combining With Perfect Phase and Zero Pointing Error} \label{Synchro}

	\subsubsection{Maximal Ratio Combiner for $\mathcal{M}$-Ary PPM} \label{MRC}

	For perfect phase synchronization, the combined signal intensity projected on the $m$th element of the array is given by (see \eqref{intensity1})
	\begin{align}
	&\lambda_s(x,y)  = \left( \sum_{i = 0}^{N-1} \| E_s^{(i)}(x,y)\|  e^{-j 2\pi(u_i x + v_i y)} \cdot \bm{1}_{A_m}(x,y)\right) \nonumber \\
	&\times \left( \sum_{\ell = 0}^{N-1} \| E_s^{(\ell)}(x,y)\|  e^{-j 2\pi(u_\ell x + v_\ell y)} \cdot \bm{1}_{A_m}(x,y)\right)^*\nonumber \\
	& = \!\! \sum_{i=0}^{N-1}\sum_{\ell=0}^{N-1}\! \|E_s^{(i)}(x,y)\| \|E_s^{(\ell)}(x,y)\|  e^{-j 2\pi((u_i-u_\ell) x + (v_i-v_\ell) y)}\nonumber \\
	&\times \bm{1}_{A_m}(x,y),
	\end{align}
	and 
	\begin{align}\label{22}
	 &\iint_{A_m}\lambda_s(x,y) \, dx \, dy \nonumber \\	&=  \sum_{i=0}^{N-1}\sum_{{\ell=0} }^{N-1} \iint_{A_m} \frac{\lambda_p}{ \rho^2} e^{-\left(\frac{x^2 + y^2}{2\rho^2}\right)} e^{-j 2\pi((u_i-u_\ell) x + (v_i-v_\ell) y)}  dx  dy \nonumber\\ 
	 &\times \|h_i\| \|h_{\ell}\| \sqrt{\beta_i \beta_\ell}. \end{align}
	Let us denote 
	\begin{align}
	x_{i,\ell}^{(m)} \triangleq \iint_{A_m} \frac{\lambda_p}{ \rho^2} e^{-\left(\frac{x^2 + y^2}{2\rho^2}\right)} e^{-j 2\pi((u_i-u_\ell) x + (v_i-v_\ell) y)} \, dx \, dy. \label{23}
	\end{align}
	Then, 
	\begin{align}
	\iint_{A_m}\lambda_s(x,y) \, dx \, dy=  \sum_{i=0}^{N-1}\sum_{{\ell=0} }^{N-1} x_{i,\ell}^{(m)}   \|h_i\| \|h_{\ell}\| \sqrt{\beta_i \beta_\ell}.
	\end{align}
	
	Let the $M \times 1$ observation vector $\by$ be denoted by $\by_s$ when the signal pulse is present, and $\by$ be denoted by $\by_n$ when only noise is present (no signal scenario). Then, for the $m$th element of the vector $\by_s$, we have that  
	\begin{align}
	Y_s^{(m)} = \sum_{i=0}^{N-1}\sum_{{\ell=0} }^{N-1} x_{i,\ell}^{(m)}   \|h_i\| \|h_{\ell}\| \sqrt{\beta_i \beta_\ell}  + V^{(m)} 
	\end{align}
	where $V^{(m)} \sim \mathcal{N}(0, \sigma^2).$
Then, for an equiprobable $\mathcal{M}$-PPM scheme, it can be shown easily that,
	\begin{align}\label{16}
&P(e|\bh) = 1 -  \left(  Q\left( -  \sqrt{  \frac{\mathcal{S}_1 }{ 2\sigma^2  }  }    \right)     \right)^{\mathcal{M}-1}, 
\end{align}
	where 
	\begin{align}
	    \mathcal{S}_1 \triangleq \sum_{m=0}^{M-1}  \left( \sum_{i=0}^{N-1}\sum_{{\ell=0} }^{N-1}  x_{i,\ell}^{(m)}   \|h_i\| \|h_{\ell}\| \sqrt{\beta_i \beta_\ell}     \right)^2.
	\end{align}

	\subsubsection{Equal Gain Combiner for $\mathcal{M}$-Ary PPM (No Channel Tap Estimation)}
	For an equal gain combiner, it can be shown (by using the same arguments put forth in Section~\ref{EGC1}) that
	\begin{align}\label{5}
	&P(e|\bh)= 1  - \left( Q \left( -  \frac{ \mathcal{S}_2     }{\sqrt{2 \sigma^2 M}}   \right)   \right)^{\mathcal{M}-1}, 
	\end{align}
	where
	\begin{align}
	    \mathcal{S}_2 \triangleq \sum_{m=0}^{M-1}  \sum_{i=0}^{N-1}\sum_{{\ell=0} }^{N-1} x_{i,\ell}^{(m)}   \|h_i\| \|h_{\ell}\| \sqrt{\beta_i \beta_\ell}. 
	\end{align}
	
	We note that when $M=1$, the probabilities of error for the MRC and EGC are equal.
	
	\textbf{Note}: Since there is a separate tracking channel for every beam in case a single detector array is used, we introduce the factor $0<\gamma<1$ that defines the fraction of energy that is received in the data channel. This implies that $1-\gamma$ of the total energy goes into the phase and beam position tracking channels. Thus, we rewrite \eqref{23} as
	\begin{align}
	x_{i,\ell}^{(m)} \triangleq \gamma \iint_{A_m} \frac{\lambda_p}{ \rho^2} e^{-\left(\frac{x^2 + y^2}{2\rho^2}\right)} e^{-j 2\pi((u_i-u_\ell) x + (v_i-v_\ell) y)} \, dx \, dy. \label{xilm2}
	\end{align}
	We use this modified expression for $x_{i,\ell}^{(m)}$ in \eqref{16} and \eqref{5}.

	\section{Beam Combining With Phase and Pointing Errors for a Single Detector Array} \label{Error}
	
	\subsection{Beam Combining With No Beam Alignment} \label{Spatial_Sync}
	In this case, we assume that none of the beams overlap on the single detector array. 
	
	\subsubsection{Maximal Ratio Combiner}
	Intuitively, we can see that the probability of error performance of the MRC in this case is the same as the performance of the MRC obtained with $N$ detector arrays. The probability of error is 
	\begin{align}
	{P(e|\bh)  =  1- \left(  Q \left(- \sqrt{  \frac{\sum_{i=0}^{N-1} \sum_{m=0}^{M-1}  \left( \| h_i\|^2 \beta_i x_i^{(m)} \right)^2}{2\sigma^2  } }   \right)      \right)^{\mathcal{M}-1}}. \label{20}
	\end{align}
	
	\subsubsection{Equal Gain Combiner}
	Since the beams do not interfere, we have that 
	$\bZ_s\sim \mathcal{N}\left( \sum_{i=0}^{N-1} \sum_{m=0}^{M-1} x_i^{(m)}, M \sigma^2  \right)$, and $\bZ_n \sim \mathcal{N}\left(0, M \sigma^2\right)$. Thus, 
	\begin{align}
	{ P(e|\bh) = 1 - \left(  Q \left(   - \frac{\sum_{i=0}^{N-1}  \sum_{m=0}^{M-1} \|h_i\|^2 \beta_i x_i^{(m)}  }{\sqrt{2\sigma^2 M}}     \right)    \right)^{\mathcal{M}-1}}. \label{21}
	\end{align}
	
	Note: The quantity $x_i^{(m)}$ in \eqref{20} and \eqref{21} is defined in \eqref{18}.

	\subsection{ Phase Synchronization Error With Zero Pointing Error} \label{pse}
	\subsubsection{Maximal Ratio Combiner}
	We assume that the $N$ beams are combined by a phase synchronization system that introduces (a small) phase error between $N$ beams. The phase error of each beam with respect to 0 radians is modeled by a zero mean Gaussian random variable with common variance $\sigma_\phi^2$. We denote these phase errors  by $\phi_0, \phi_1, \dotsc, \phi_{N-1}$. Additionally, we assume that all the phase errors are independent of each other, i.e., $\phi_i \perp \phi_j$ for $i \neq j$.

	 Let us denote the electric field of the $n$th beam by $E_s^{(n)}(x,y)$. After phase correction and perfect beam alignment, the resultant electric field on the array is given by
	 \begin{align}
	 \mathcal{E}(x, y) \triangleq \sum_{i=0}^{N-1} \|E_s^{(i)}(x,y)\| e^{j \phi_i} e^{-j 2\pi(u_i x + v_i y)} \cdot \mathbbm{1}_\mathcal{A}(x,y).
	 \end{align}
 
	By using the same arguments as in Section~\ref{MRC}, we have that the probability of error for the maximal ratio combiner is,
	 \begin{align} \label{24}
	& P(e|\bh) = 1 - \left( Q \left( - \sqrt{  \frac{\mathcal{S}_3 }{ 2\sigma^2 }  }     \right)      \right)^{\mathcal{M}-1},
	 \end{align}
	 where
	 \begin{align}
	     \mathcal{S}_3 \triangleq \sum_{m=0}^{M-1} \left(    \sum_{i=0}^{N-1} \sum_{\ell=0}^{N-1} x_{i,\ell}^{(m)}  \|h_i\| \|h_\ell\| \sqrt{\beta_i \beta_\ell} e^{j(\phi_i -\phi_\ell)}  \right)^2.
	 \end{align}

	 \subsubsection{Equal Gain Combiner}
	 It is straightforward to show that the probability of error for the equal gain combiner is 
	 \begin{align} \label{25}
	 & P(e|\bh) = 1 -  \left(   Q  \left( -   \frac{ \mathcal{S}_4}{ \sqrt{2\sigma^2 M}    }       \right)   \right)^{\mathcal{M}-1},
	 \end{align}
	 where 
	 \begin{align}
	     \mathcal{S}_4 \triangleq \sum_{m=0}^{M-1}   \sum_{i=0}^{N-1} \sum_{\ell=0}^{N-1} x_{i,\ell}^{(m)}  \|h_i\| \|h_\ell\| \sqrt{\beta_i \beta_\ell} e^{j(\phi_i -\phi_\ell)}.   
	 \end{align}
	 
	 Note: The quantity $x_{i,\ell}^{(m)}$ in \eqref{24} and \eqref{25} is defined in \eqref{xilm2}.

	 \subsubsection{Large $N$ and Uniform Phase Error Case} \label{asymptotic}
	 The probability of error expressions in \eqref{MRC_Pe}, \eqref{EGC_Pe}, \eqref{24} and \eqref{25} contain sums that are functions of $N$, and the complexity of these expressions grows with $N$. In order to analyze the effect of large $N$ and large $M$ on the probability of error, we use an approximate technique which does not depend on $N$ for the purpose of computation, and thus the complexity is only dependent on the number of detectors $M$. This technique is discussed in the remainder of this section.
	 
	  Let us assume that we have a large number of beams that lie in close proximity to the center of the beam combining lens, i.e., $u_i \approx v_i \approx 0$ for all $i$. The assumption concerning the close proximity to the center of the lens is made in order to simplify the analysis. Let the quantity $\lambda_s(x,y) \triangleq  \frac{\lambda_p}{\rho^2 } \exp\left(  - \frac{x^2 + y^2}{2 \rho^2}    \right) \cdot \bm{1}_\mathcal{A}(x,y)$ correspond to the total transmitted signal intensity.  By Euler expansion, we have that the total electric field on the array is
	 \begin{align}\label{4}
	 &\mathcal{E}(x,y) =  \sum_{i=0}^{N-1}\|E_s^{(i)}(x,y)\| e^{j\phi_i} e^{-j 2\pi(u_i x + v_i y)} \cdot \bm{1}_{\mathcal{A}}(x,y) \nonumber \\
	 & = \sum_{i=0}^{N-1}\|E_s^{(i)}(x,y)\| \cos(\phi_i-2\pi(u_ix+v_iy))\cdot \bm{1}_{\mathcal{A}}(x,y) \nonumber \\
	 &+ j \sum_{i=0}^{N-1}\|E_s^{(i)}(x,y)\| \sin(\phi_i-2\pi(u_ix+v_iy))\cdot \bm{1}_{\mathcal{A}}(x,y) \nonumber \\
	 & \approx \sqrt{\frac{{\lambda_p} }{\rho^2 } e^{  - \frac{\left(x^2 + y^2\right)}{2 \rho^2}     } } \nonumber \\
	 & \times \!\left(  \sum_{i=0}^{N-1} \|h_i\| \sqrt{\beta_i} \cos(\phi_i) + j\! \sum_{i=0}^{N-1} \! \|h_i\| \sqrt{\beta_i} \sin(\phi_i) \!\right)\!\cdot \!\bm{1}_{\mathcal{A}}(x,y),
	 \end{align}
	where the approximation follows since $u_i \approx v_i \approx 0$ for all $i$. Let us denote $\|h_i\| \sqrt{\beta_i}$ by $\tilde{h}_i$. We assume that $\phi_i \sim \mathcal{U}(-\pi, \pi)$ for all $i$. Moreover, $ \tilde{h}_i   \perp \phi_i$, $\tilde{ h}_n  \perp \phi_i$,  and $\phi_i \perp \phi_{\ell}$ for $i \neq \ell$, where the symbol $\perp$ is used to indicate statistical independence. Let us denote the real part by $X$: $X \triangleq \sum_{i=0}^{N-1}\tilde{h}_i \cos(\phi_i)$, and the imaginary part by $Y$: $Y \triangleq \sum_{n=0}^{N-1}\tilde{h}_i \sin(\phi_i)$. The sum intensity in this case becomes
	 \begin{align}
	 &\lambda(x,y) = \frac{{\lambda_p} }{\rho^2 } e^{  - \frac{\left(x^2 + y^2\right)}{2 \rho^2} }  \| X + jY \|^2 \cdot \bm{1}_{\mathcal{A}}(x,y) \nonumber \\
	 &= \frac{{\lambda_p} }{\rho^2 } e^{  - \frac{\left(x^2 + y^2\right)}{2 \rho^2}  } \left( X^2 + Y^2 \right)\cdot \bm{1}_{\mathcal{A}}(x,y).
	 \end{align}
	 For large $N$, both $X$ and $Y$ converge in distribution to   Gaussian random variables via the \emph{Central Limit Theorem}. It can be shown easily that $\E[X] = \E[Y] = 0$, and $\E[XY] = 0$: thus $X$ and $Y$ are uncorrelated Gaussian random variables. Additionally, 
	 \begin{align}
	 &\var[X]  = \E[X^2] = \E\left[  \sum_{i=0}^{N-1} \sum_{\ell=0}^{N-1} \tilde{ h}_i  \tilde{h}_\ell  \cos(\phi_i) \cos(\phi_\ell)   \right] \nonumber \\
	 &= \sum_{i=0}^{N-1} \E \left[ \tilde{h}_i^2   \right] \E[\cos^2(\phi_i)].
	 \end{align}
	  It can be shown easily that $\E[\cos^2(\phi_i)] = \frac{1}{2}$. Hence,
	 \begin{align}
	 \var[X] = \frac{1}{2} \sum_{i=0}^{N-1} \E \left[ \tilde{h}_i^2   \right].
	 \end{align}
	 Similarly, we can show that $\var[Y] = \var[X] = \frac{1}{2} \sum_{i=0}^{N-1} \E \left[ \tilde{h}_i^2   \right].$ Considering \eqref{19}, we note that $\tilde{h}_i = \sqrt{P_t} \frac{\|h_i\|^2}{\sqrt{\sum_{j = 0}^{N-1}\|h_j\|^2}}.$ Intuitively, it is easy to see that the $\tilde{h}_i$ are identically distributed with $\E[\tilde{h}_i^2] =\sigma_h^2$ for some real number $\sigma_h^2 > 0$. Therefore, $\var[X] = \frac{N}{2} \sigma_h^2$.
	 
	 We now have that $W \triangleq (X^2 + Y^2)$ is an exponential random variable with parameter 
	 \begin{align}
	 \frac{1}{2 \var[X]} = \frac{1}{N \sigma_h^2}. \label{7}
	 \end{align}
	 Thus, the peak of the resulting intensity (located at $(x_0, y_0)$) fluctuates according to the exponential distribution (with parameter given in \eqref{7}) when the number of beams is large and they lie in close proximity to each other. In this case, the conditional probability of error for the MRC can be shown to be
	 \begin{align}
	 	P\left(e|W=w\right) = 1 - \left(   Q\left( - w\sqrt{  \frac{\sum_{m=0}^{M-1}  \left(  x^{(m)}     \right)^2 }{ 2\sigma^2  }  }     \right)      \right)^{\mathcal{M}-1}, \label{17}
	 \end{align}
	 where $x^{(m)} \triangleq \iint_{A_m} \frac{{\lambda_p} }{\rho^2 } e^{  - \frac{\left(x^2 + y^2\right)}{2 \rho^2} } \, dx \, dy$. The probability of error is obtained by $P(e) = \int_{-\infty}^\infty P\left(e|W=w\right)f_W(w)\, dw$. In a similar fashion, the conditional  probability of error for the EGC can be shown to be
	 \begin{align}
	 P\left(e|W=w\right) = 1 - \left(   Q \left( -  w\frac{ \sum_{m=0}^{M-1}  x^{(m)}   }{\sqrt{2\sigma^2 M}}   \right)   \right)^{\mathcal{M}-1}. \label{8}
	 \end{align}

	 \subsection{Pointing Error Effect With Perfect Phase Synchronization} \label{Phase_Sync}
	 In this case, we assume that the beam alignment is not perfect due to nonzero pointing error. This means that there is at least one beam that leaves the coherence region $\mathcal{A}_c$ due to random disturbances every now and then before the tracking assembly can bring the beam back inside $\mathcal{A}_c$. We assume that if two or more beams leave $\mathcal{A}_c$, they do not interfere with each other in the region $\mathcal{A} - \mathcal{A}_c$.
	 
	 Let the set $\mathcal{B}$ contain the indices of the beams that lie in $\mathcal{A}_c$, and $\mathcal{B}'$ is the set of indices pertaining to the beams that do not lie in $\mathcal{A}_c$.
	 In such a case, the output of the $m$th cell is 
	 \begin{align}
	 Y_s^{(m)} = x^{(m)} \left( \sum_{i \in \mathcal{B}}   \|h_i\| \sqrt{\beta_i}  \right)^2 + \sum_{j \in \mathcal{B}'} x_j^{(m)}  \| h_j\|^2 \beta_j + V^{(m)}.
	 \end{align}
	 
	 The evolution of $\bmx_n$ in terms of $\bmx_0$ is furnished by \eqref{evo}. Let us assume that $n \to \infty$ so that steady state is achieved by the system, and $  \bmx \triangleq \lim\limits_{n \to \infty} \bmx_n$. Then, the probability that there are $N_0 < N$ beam outside the coherence region  is 
	 \begin{align}
	P(\{ N_0\text{ beams in } \mathcal{A}_c'  \}) = \left( P\left( \left\{   \| \bmx\| > \rho   \right\}   \right)\right)^{N_0}.
	 \end{align}
	 For the simplest possible case let us assume that $\bP$ is diagonal with equal eigenvalues $a$, where $|a| < 1$. When the diagonal values are equal, then the variance of each dimension of $\bmx$ would also be the same. Additionally, let us denote the variance of each of the two dimensions of $\bmx$ by  $\sigma_x^2$. Then, from \eqref{6}, we have that when $n$ is large,
	\begin{align}
	\sigma_x^2 = \lim\limits_{n \to \infty} \sigma_W^2 \sum_{j=0}^{n}a^{2j} = \sigma_W^2 \left( \frac{1}{1-a^2} \right).
	\end{align}
	Then,  it can be shown easily that 
	\begin{align}
	P\left( \left\{   \| \bmx\| > \rho   \right\}   \right) = \exp\left( \frac{-\rho^2}{2\sigma_x^2}\right).
	\end{align}
	The number of beams $N_0$ in $\mathcal{B}'$ is modeled by the \emph{Binomial distribution}:
	\begin{align}
	    &P(\{N_0 = n_0\}) = {N \choose n_0}  \left(\exp\left(- \frac{\rho^2}{2\sigma_x^2}\right)\right)^{n_0}\nonumber \\
	    & \times \left(1 - \exp\left( -\frac{\rho^2}{2\sigma_x^2}\right) \right)^{N-n_0}.
	\end{align}
	  
	 \subsubsection{Maximal Ratio Combiner}
	  For the maximal ratio combiner, the probability of error is given by 
	  \begin{align}
	& P(e|\bh)  =  \sum_{n_0 =0}^{N}P(e| N_0 = n_0) P(\{ N_0=n_0\}) \nonumber \\
	 & = \sum_{n_0 =0}^{N} \left[1 - \left(   Q\left( - \sqrt{  \frac{\mathbb{S}_1 }{ 2\sigma^2  }  }     \right)      \right)^{\mathcal{M}-1 } \right] \nonumber \\
	 & \times {N \choose n_0}  \left(\exp\left(- \frac{\rho^2}{2\sigma_x^2}\right)\right)^{n_0} \left(1 - \exp\left( -\frac{\rho^2}{2\sigma_x^2}\right) \right)^{N-n_0}, \label{11}
	 \end{align}
	 where 
	 \begin{align}
	     \mathbb{S}_1 \! \triangleq \! \sum_{m=0}^{M-1}\! \left(  \!
	     \sum_{i \in \mathcal{B}} \sum_{\ell \in \mathcal{B}} x_{i,\ell}^{(m)}  \|h_i\| \|h_\ell\| \sqrt{\beta_i\beta_\ell}   + \!\sum_{j \in \mathcal{B}'} x_j^{(m)} \| h_j\|^2 \beta_j \! \right)^{\!\!2}\!\!.
	 \end{align}
	 We note that there is no overlap between any beam in $ \mathcal{B}'$. Additionally, we note that there is no overlap between any beam in  $\mathcal{B}$ and any beam in $ \mathcal{B}'$. Thus, 
	  \begin{align}
	& \left(   \sum_{i \in \mathcal{B}} \sum_{\ell \in \mathcal{B}} x_{i,\ell}^{(m)}  \|h_i\| \|h_\ell\| \sqrt{\beta_i\beta_\ell}   + \sum_{j \in \mathcal{B}'} x_j^{(m)} \| h_j\|^2 \beta_j \right)^2 \nonumber \\
	&=  \left(   \sum_{i \in \mathcal{B}} \sum_{\ell \in \mathcal{B}} x_{i,\ell}^{(m)}  \|h_i\| \|h_\ell\| \sqrt{\beta_i\beta_\ell} \right)^2 + \left(\sum_{j \in \mathcal{B}'} x_j^{(m)} \| h_j\|^2 \beta_j \right)^2    \nonumber \\
	 & = \left(   \sum_{i \in \mathcal{B}} \sum_{\ell \in \mathcal{B}} x_{i,\ell}^{(m)}  \|h_i\| \|h_\ell\| \sqrt{\beta_i\beta_\ell} \right)^2 + \sum_{j \in \mathcal{B}'} \left(x_j^{(m)} \| h_j\|^2 \beta_j \right)^2 
	 \end{align}
	Thus, we can write \eqref{11} more simply as
	
	 \begin{align}
	 &P(e|\bh) = \sum_{n_0 =0}^{N-1} \left[1 - \left(   Q\left( - \sqrt{  \frac{\mathbb{S}_2 }{ 2\sigma^2  }  }     \right)      \right)^{\mathcal{M}-1 } \right] \nonumber \\
	 & \times {N \choose n_0}\left(\exp\left(- \frac{\rho^2}{2\sigma_x^2}\right)\right)^{n_0} 
	 	 \left(1 - \exp\left( -\frac{\rho^2}{2\sigma_x^2}\right) \right)^{N-n_0}. \label{12}
	 \end{align}
	 where 
	 \begin{align}
	     \mathbb{S}_2 &\triangleq \sum_{m=0}^{M-1}  \left(   \sum_{i \in \mathcal{B}} \sum_{\ell \in \mathcal{B}} x_{i,\ell}^{(m)}  \|h_i\| \|h_\ell\| \sqrt{\beta_i\beta_\ell} \right)^2 \nonumber \\ 
	     &+ \sum_{m=0}^{M-1}\sum_{j \in \mathcal{B}'} \left(x_j^{(m)} \| h_j\|^2 \beta_j  \right)^2. 
	 \end{align}
	  
	 \subsubsection{Equal Gain Combiner}
	 For the equal gain combiner case, it can be shown that the probability of error is 
	 \begin{align}
	 &P(e|\bh) = \sum_{n_0 =0}^{N-1} \left[1 - \left(   Q\left( - {  \frac{\mathbb{S}_3 }  { \sqrt{2\sigma^2 M}  }  }     \right)      \right)^{\mathcal{M}-1 } \right]  \nonumber \\
	 &\times {N \choose n_0} \left(\exp\left(- \frac{\rho^2}{2\sigma_x^2}\right)\right)^{n_0}  \left(1 - \exp\left( -\frac{\rho^2}{2\sigma_x^2}\right) \right)^{N-n_0}, \label{13}
	 \end{align}
	 where
	 \begin{align}
	     \mathbb{S}_3 & \triangleq \sum_{m=0}^{M-1}    \sum_{i \in \mathcal{B}} \sum_{\ell \in \mathcal{B}} x_{i,\ell}^{(m)}  \|h_i\| \|h_\ell\| \sqrt{\beta_i \beta_\ell} \nonumber \\
	     &+ \sum_{m=0}^{M-1} \sum_{j \in \mathcal{B}'} x_j^{(m)} \| h_j\|^2 \beta_j. 
	 \end{align}
	
	 \subsection{Pointing Error Effect with Imperfect Phase Synchronization}
	 Incorporating the effects of both imperfect phase and nonzero pointing errors, we have the following expressions for the MRC and EGC receivers.
	 \subsubsection{Maximal Ratio Combiner}
	 The probability of error for maximal ratio combiner is
	 \begin{align}
	 &P(e|\bh) = \sum_{n_0 =0}^{N-1} \left[1 - \left(   Q\left( - \sqrt{  \frac{\mathbb{S}_4 }{ 2\sigma^2  }  }     \right)      \right)^{\mathcal{M}-1 } \right] \nonumber \\
	 & \times {N \choose n_0}\left(\exp\left(- \frac{\rho^2}{2\sigma_x^2}\right)\right)^{n_0} 
	 \left(1 - \exp\left( -\frac{\rho^2}{2\sigma_x^2}\right) \right)^{N-n_0}, \label{26}
	 \end{align}
	 where 
	 \begin{align}
	     \mathbb{S}_4 & \triangleq \sum_{m=0}^{M-1}  \left(   \sum_{i \in \mathcal{B}} \sum_{\ell \in \mathcal{B}} x_{i,\ell}^{(m)}  \|h_i\| \|h_\ell\| \sqrt{\beta_i\beta_\ell} \,e^{j(\phi_i-\phi_\ell)} \right)^2 \nonumber \\
	     & + \sum_{m=0}^{M-1}\sum_{j \in \mathcal{B}'} \left(x_j^{(m)} \| h_j\|^2 \beta_j  \right)^2. 
	 \end{align}
	 
	 \subsubsection{Equal Gain Combiner}
	 The probability of error for equal gain combiner is
	\begin{align}
	&P(e|\bh) = \sum_{n_0 =0}^{N-1} \left[1 - \left(   Q\left( - {  \frac{\mathbb{S}_5 }  { \sqrt{2\sigma^2 M}  }  }     \right)      \right)^{\mathcal{M}-1 } \right]  \nonumber \\
	&\times {N \choose n_0} \left(\exp\left(- \frac{\rho^2}{2\sigma_x^2}\right)\right)^{n_0}  \left(1 - \exp\left( -\frac{\rho^2}{2\sigma_x^2}\right) \right)^{N-n_0}, \label{15}
	\end{align}
	where
	\begin{align}
	    \mathbb{S}_5 & \triangleq \sum_{m=0}^{M-1}    \sum_{i \in \mathcal{B}} \sum_{\ell \in \mathcal{B}} x_{i,\ell}^{(m)}  \|h_i\| \|h_\ell\| \sqrt{\beta_i \beta_\ell}\, e^{j(\phi_i-\phi_\ell)} \nonumber \\
	    & + \sum_{m=0}^{M-1} \sum_{j \in \mathcal{B}'} x_j^{(m)} \| h_j\|^2 \beta_j. 
	\end{align}
	
	\section{Optimization of Bit Error Rate in Terms of Beam Radius for a Single Detector Array} \label{Optimize}
	As seen in \eqref{26} and \eqref{15}, the probability of error is a function of both beam radius $\rho$ and  pointing error variance $\sigma_x^2$. It is more advantageous to keep the centers of the beams aligned on the detector array and let the beams overlap if the phase error between the beams is below a certain threshold. This is because in case the phase errors are small, the beams will add (partially) coherently over a certain (small) number of detectors, and the output from other detectors in the array can be ignored. This will minimize the total noise from the sufficient statistic which will improve the resulting SNR and the probability of error. 
	
	However, if the beam radius is too small, the probability that the beams will overlap---and interfere partially coherently---will become smaller. This will diminish the resulting SNR and result in a higher probability of error.
	
	Thus, in terms of optimizing the beam radius in order to minimize the probability of error, we have the following minimization problem at our hands:
	\begin{equation*}
\begin{aligned}
& \underset{\rho}{\text{minimize}}
& &  P(e) \\
& \text{subject to}
& & i) \, \rho_{\text{min}} < \rho < \rho_{\text{max}},\\
& & & ii)\, \textsf{SNR} = S_0, \\
& & & iii)\, \sigma_x = \sigma_0,\\
& & & iv) \, N = N_0, \\
& & & v) \, M = M_0.
\end{aligned}
\end{equation*}
In this optimization problem, $S_0$, $\sigma_0$, $N_0$ and $M_0$ are constants. Additionally, since \eqref{26} and \eqref{15} are both highly nonlinear in terms of $\rho$, we resort to a global optimization routine such as a \emph{real number genetic algorithm} in order to minimize the probability of error. The readers may see \cite{Rao} for a more detailed discussion on real number genetic algorithms.

	\section{Experimental Results} \label{Experiments}
	\begin{figure}[H]
		\centering
		\begin{tabular}{ll}
			\includegraphics[scale=0.9]{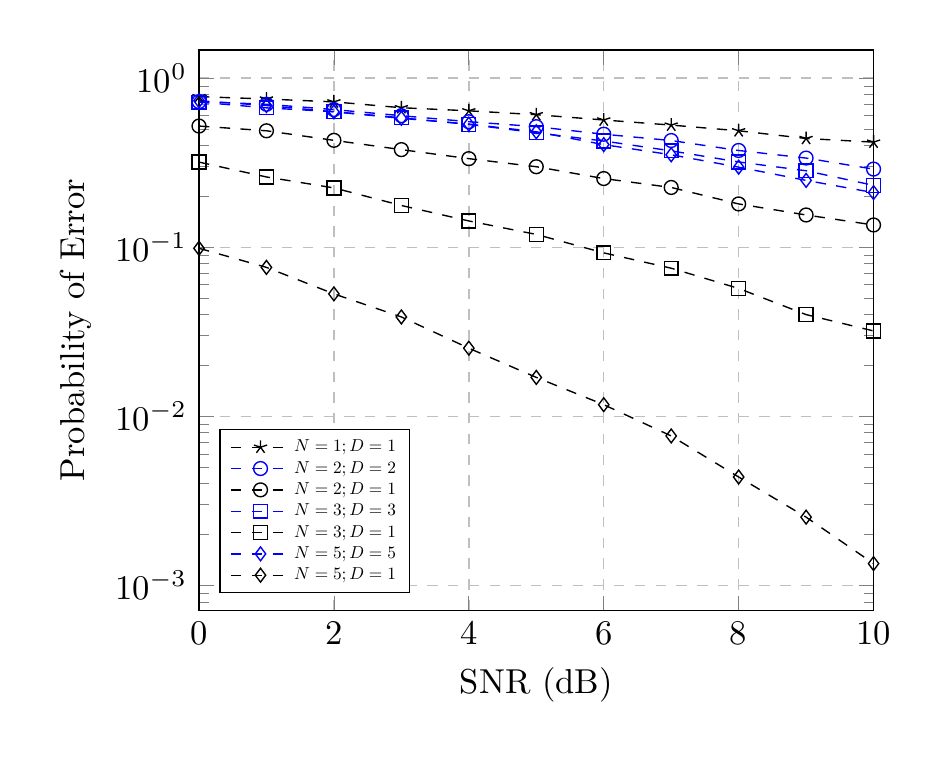}
			\\
			\includegraphics[scale=0.9]{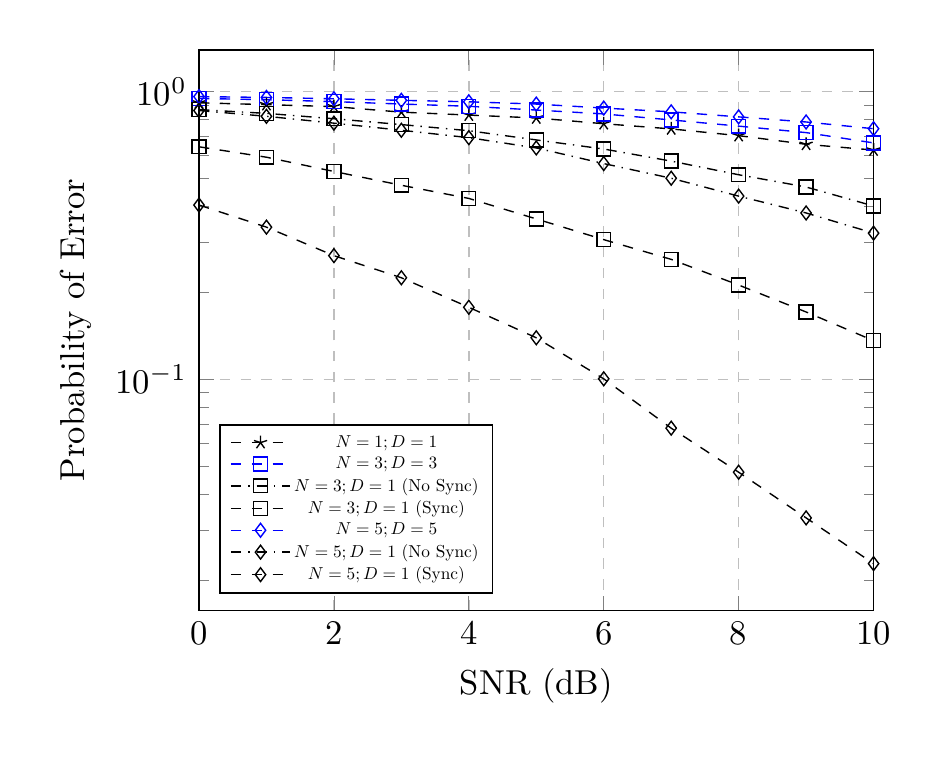}
		\end{tabular}
	\caption{Probability of error as a function of SNR for the  maximal ratio combining (top) and equal gain combining (bottom) cases with perfect phase synchronization and zero pointing error.  The number of detectors in each array is $M=16$, and the beam radius is 0.2 millimeters. The area of each array is 4 square millimeters. The mean of the exponential fading distribution is 0.5. The experiments are carried out for $8$-PPM case. In the legend, the number $N$ represents the number of beams, and $D$ represents the number of detector arrays. The total received power in each of the schemes is the same.} \label{plot1}
	\end{figure}
	Fig.~\ref{plot1} presents the probability of error graphs for the MRC and EGC receivers for the perfect phase synchronization and zero pointing error case. We note that the system with one array of detectors markedly outperforms the systems with multiple arrays for both MRC and EGC.
	\begin{figure}[H]
\centering
\begin{tabular}{ll}
\includegraphics[scale=0.9]{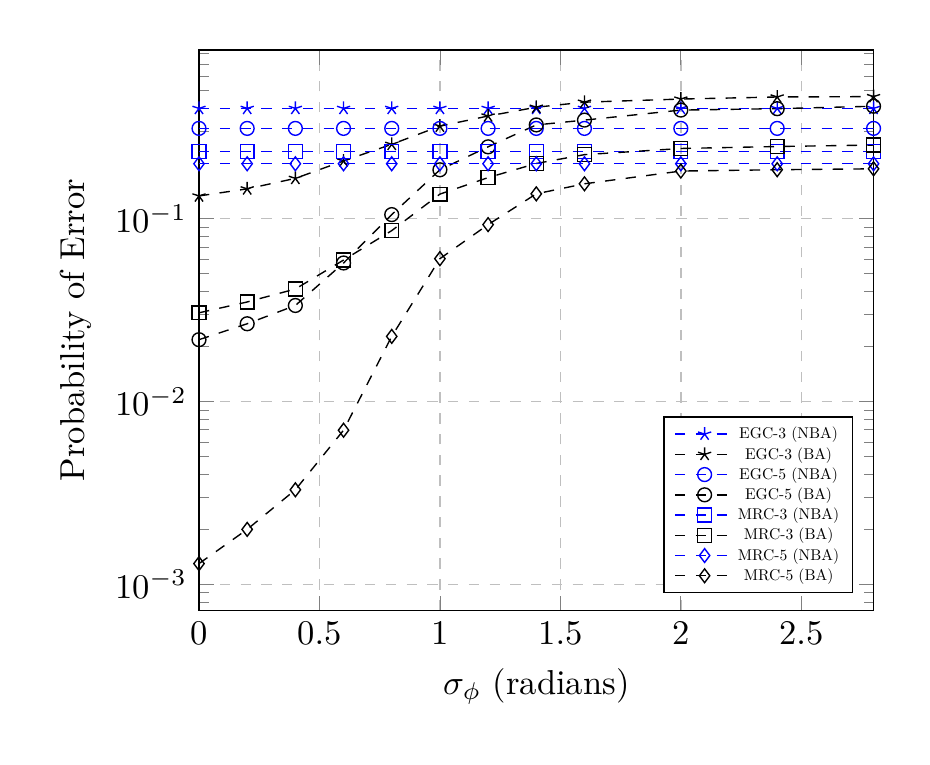} \\
\includegraphics[scale=0.9]{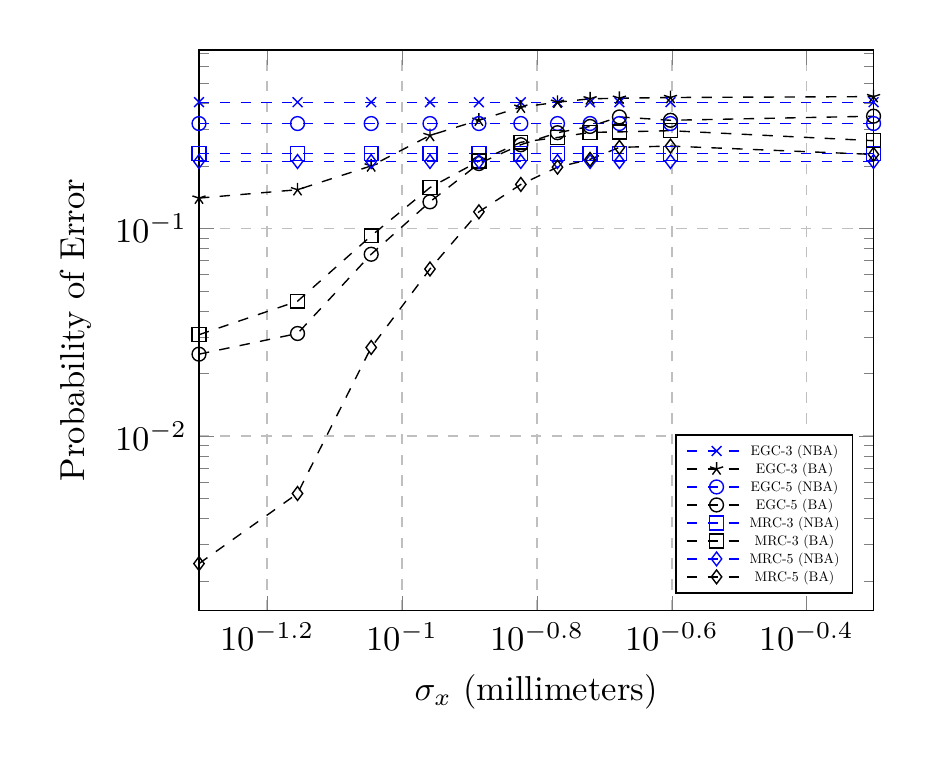}
\end{tabular}
\caption{Probability of error for different scenarios as a function of phase error $\phi$ (top) and pointing error standard deviation $\sigma_x$ (bottom) when a single detector array is used at the receiver. In the legend, the number indicates the number of beams $N$, and BA corresponds to ``beam alignment'' case, and NBA stands for ``no beam alignment'' case. The SNR is fixed at $10$ dB, the beam radius $\rho$ is set at $0.2$ millimeters, and the area of the array is $4$ square millimeters. The mean of the exponential fading is 0.5, and the experiments are carried out for the 8-PPM scenario. For the figure on the top, $\sigma_x = 0$ millimeters, and for the figure on the bottom, $\sigma_\phi =0$ radians. The total received power in each of the schemes is the same.} \label{plot2}
\end{figure}	 
Fig.~\ref{plot2} indicates the probability of error performance when phase and pointing error is introduced. For both the figures, we note that the beam alignment provides better performance when the phase and pointing errors are below certain thresholds for both the MRC and EGC receivers. Additionally, as expected, the performance of these systems converges to the no beam alignment case as the phase and pointing errors exceed the threshold values. 
\begin{figure}[H]
    \centering
    \includegraphics[scale=0.9]{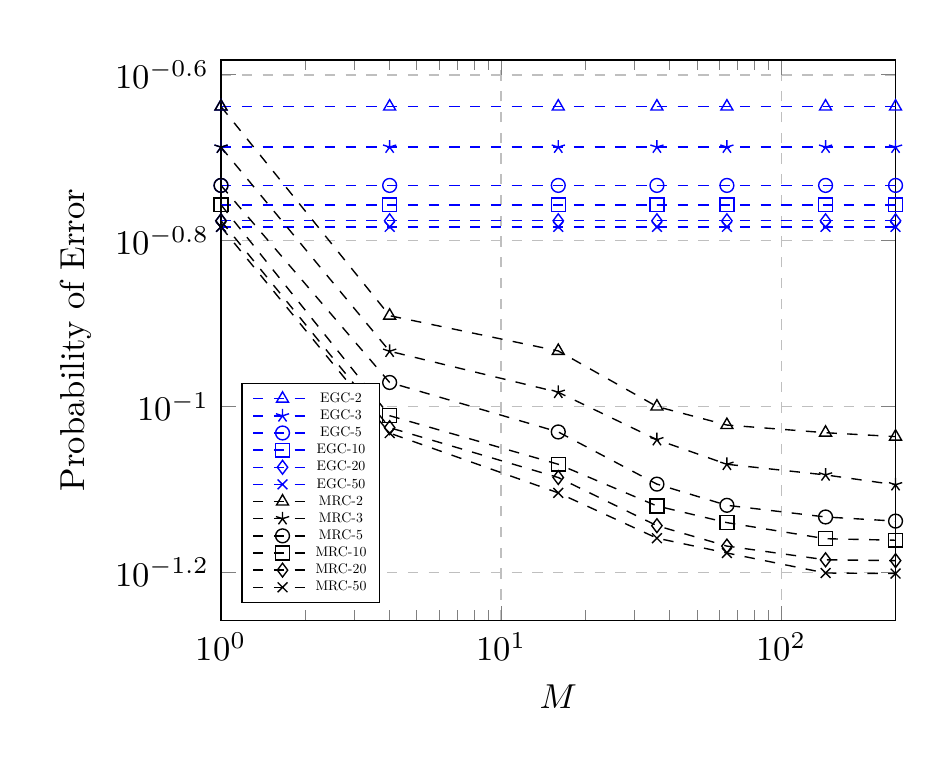}
    \caption{Probability of error as a function of number of cells $M$ for different number of beams $N$. The SNR is fixed at $20$ dB, the beam radius is set at $0.2$ millimeters, and the area of the array is $4$ square millimeters. The mean of the exponential fading is 0.5, and the experiments are carried out for the 8-PPM scenario. In the legend, the digit indicates the number of beams corresponding to a particular scheme. For example, MRC-5 corresponds to the maximal ratio combining receiver with 5 beams. The total received power in each of the schemes is the same.}
    \label{plot3}
\end{figure}	 
Fig.~\ref{plot3} presents the probability of error for the large $N$ and $M$, and uniform phase error approximation (see \eqref{17} and \eqref{8}) for the single array of detectors. We observe that the margin in improvement in the probability of error diminishes as we increase either $N$ or $M$. As expected, we also note that the performance of EGC receiver is independent of the number of detectors $M$ in the array.
\begin{figure}[H]
		\centering
		\begin{tabular}{ll}
			\includegraphics[scale=0.9]{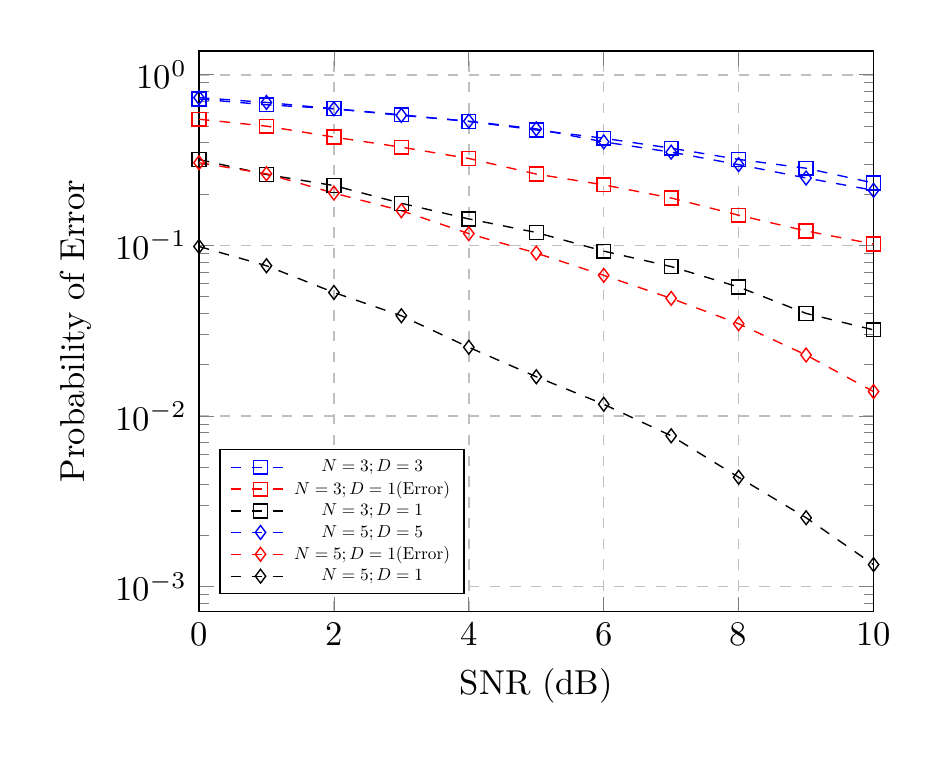}
			\\
			\includegraphics[scale=0.9]{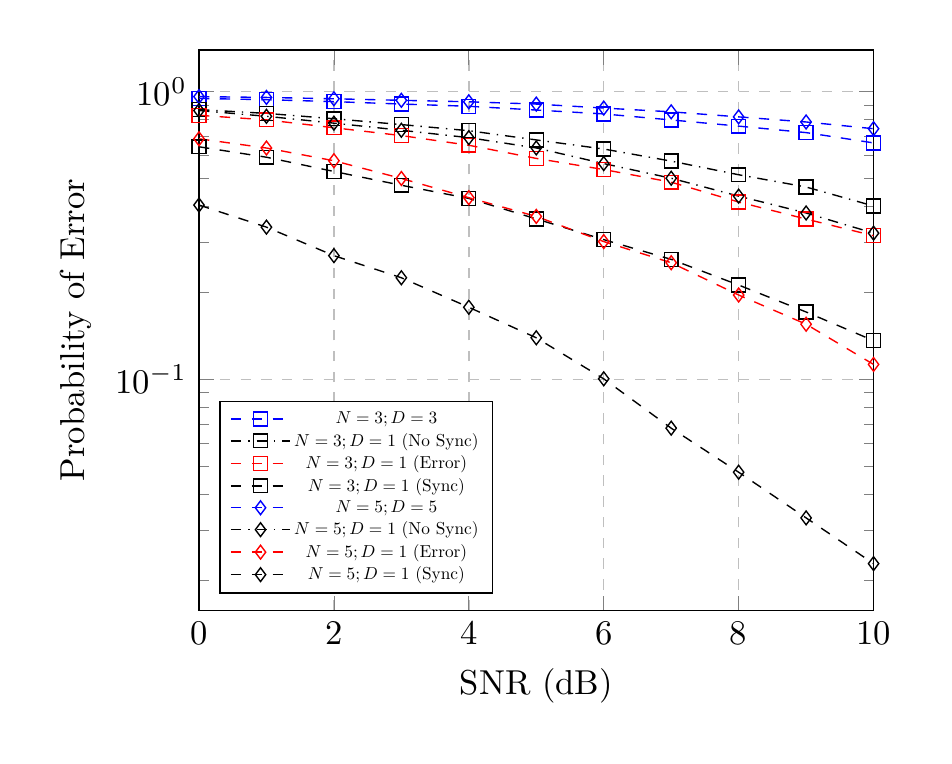}
		\end{tabular}
	\caption{Probability of error as a function of SNR for the  maximal ratio combining (top) and equal gain combining (bottom) cases when imperfect phase synchronization and nonzero pointing error is taken into account. The number of detectors in each array is $M=16$, and the beam radius is 0.2 millimeters. The area of each array is 4 square millimeters. The mean of the exponential fading distribution is 0.5, phase error standard deviation $\sigma_{\phi} = 0.5$ radians, pointing error standard deviation $\sigma_x = 0.05$ millimeters and the power split factor $\gamma = 0.7$. The experiments are carried out for $8$-PPM case. The total received power in each of the schemes is the same.} \label{plot4}
	\end{figure}
	Fig.~\ref{plot4} illustrates the probability of error for a practical MRC and EGC system that suffers from phase and pointing errors. For these plots, it is assumed that the tracking and phase correction channels take up 30 percent of the total received power, and the remaining 70 percent is utilized for detecting the PPM symbol.
	
\begin{figure}[H]
	    \centering
	    \includegraphics[scale=0.9]{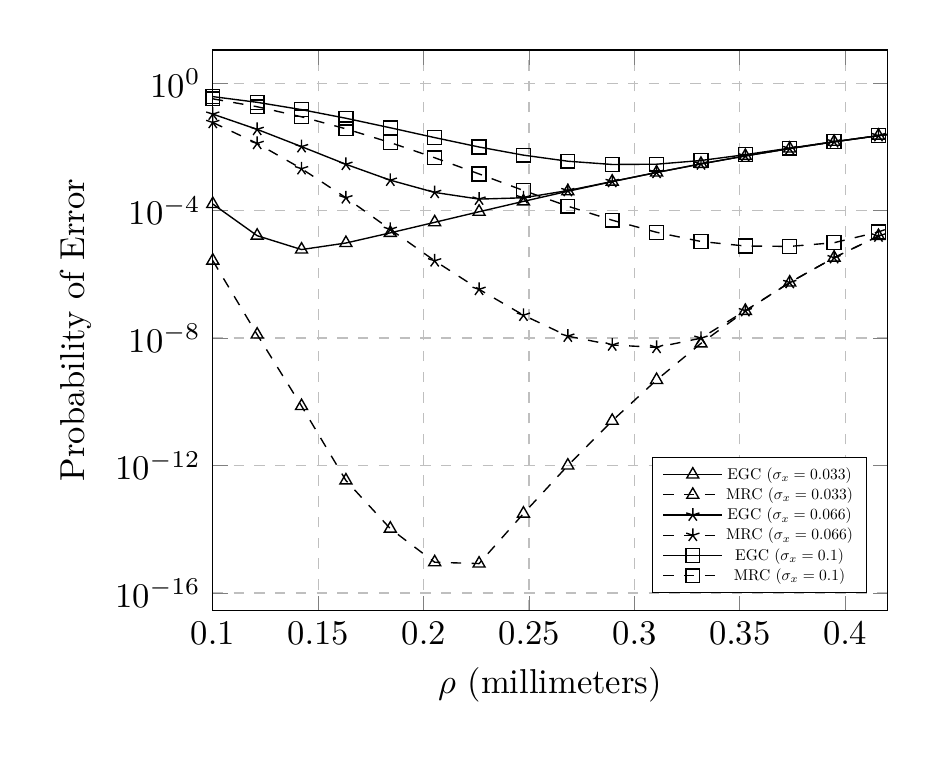}
	    \caption{Probability of error as a function of beam radius $\rho$ and three different values of pointing error standard deviation $\sigma_x$. In this case, a single detector array was employed at the receiver. The \snr  is fixed at 3 dB. The number of beams $N=5$, and the number of detectors in the array $M=16$. }
	    \label{plot5}
	\end{figure}

	Fig.~\ref{plot5} delineates the probability of error curves as a function of beam radius $\rho$ for different values of the pointing error standard deviation $\sigma_x$. This plot shows the dependence of optimal $\rho^*$ on $\sigma_x$. For instance, if $\sigma_x$ is large, we have to increase the size of beam radius in order for the overlap (and partial coherent interference) to occur frequently. However, for a large beam radius, the beam projects (same amount of) energy onto a  large number of detectors, thereby lowering the SNR of the resulting sufficient statistic\footnote{This is especially true for the MRC receiver.}. Therefore, the optimal beam radius is a trade-off between these two extremes that provides the minimum probability of error.
	
	\begin{figure}[H]
	    \centering
	    \begin{tabular}{ll}
	    \includegraphics{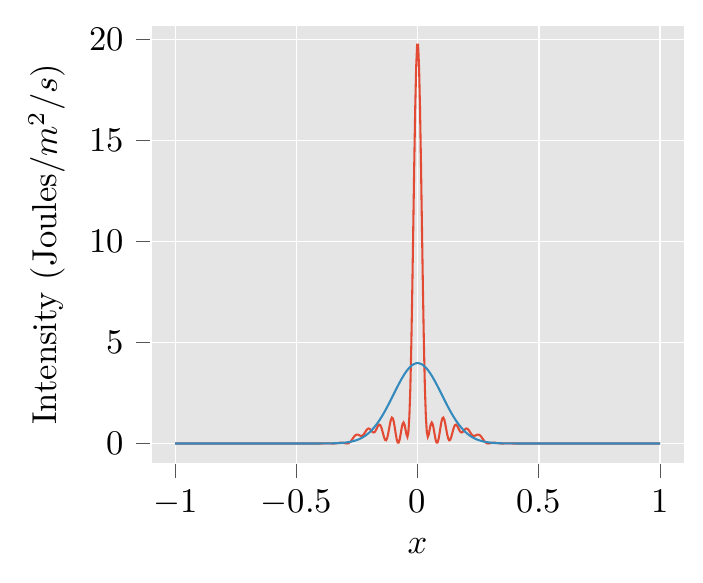} \\
	    \includegraphics{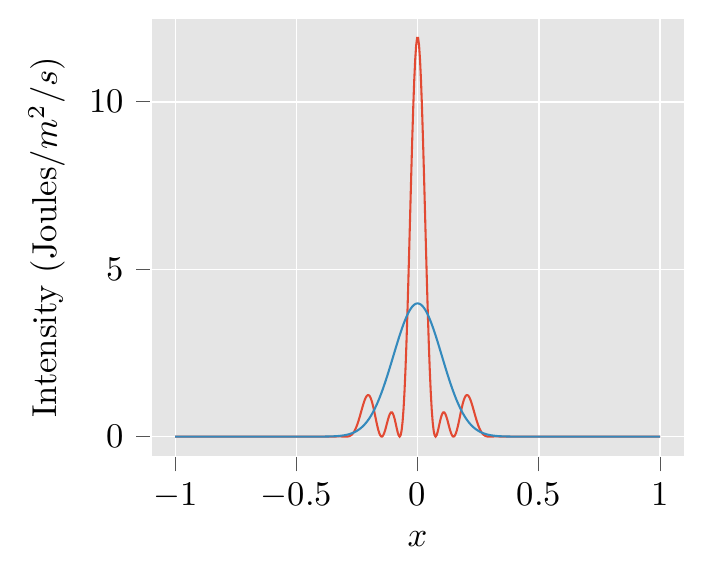}
	   \end{tabular}
	    \caption{Comparison of coherent beam combining for the $N=5$ beams (top) and $N=3$ beams (bottom) scenarios as observed along $x$-axis (one dimensional case). The graph in red is the intensity of the sum airy pattern representative of perfect coherent beam combining ($ \phi_{i}-\phi_{j} = 0$ for all $i$ and $j$). The sidelobes occur due to nonzero spatial phase error, and the magnitude of the sidelobes depends on the locations $(u_i, v_i)$ of the beams incident on the combining lens.}
	    \label{intensity_plots}
	\end{figure}
	Fig.~\ref{intensity_plots} depicts the effect of coherent beam combining on the intensity of the resulting signal. We have plotted the intensity of the resulting airy pattern along one dimension of the array for the purpose of clarity. The plot in red indicates coherent beam combining: $\exp(j (\phi_{i}-\phi_{j})) = 1$ for all $i$ and $j$, where $i$ or $j$ is the beam index. However, we observe sidelobes due to the spatial phase error along $x$-axis, $\exp\left(-j 2 \pi (u_i x)\right)$, and the magnitude of these sidelobes depends on $u_i$'s. For $N=5,$ $u_i$'s are chosen to be $0, 4, -4, -8$ and $11$, and for $N=3,$ $u_i$'s are $0, 4$ and $-4$ on a millimeter scale. In contrast, the graph in blue represents a scalar sum of the intensities of different beams. We note that coherent combining or spatial synchronization leads to higher peaks in the intensity of the resulting beam if the phase difference $ (\phi_i - \phi_j)$ between different beams is minimized. However, the radius of the resulting beam has to diminish (in comparison to the blue plot) since the total energy of the resulting beam for the blue and red plots is the same. Finally, the radius of the resulting beam after combining is a function of $N$: a larger $N$ creates a more focused  beam with a smaller beam radius.
	
	\section{Conclusion} \label{Conc}
	In this paper, we have analyzed the probability of error for a free-space optical MISO system that utilizes an array of detectors. In this regard, the probability of error was analyzed under conditions of phase and pointing errors of different beams. We conclude that a single detector array yields a lower probability of error than the multiple array scheme if the variance of phase and pointing errors can be restricted below certain thresholds. Furthermore, the probability of error of a single detector array can be further optimized by selecting an optimal beam radius that is a function of the pointing error variance. Finally, even though we have only considered MRC and EGC fusion algorithms for the array of
detectors in this paper, the lower combining complexity schemes such as \cite{Alouini:01, Simon:01, Yang:05} can be considered as part of future studies.

	\bibliography{refs.bib}

	\bibliographystyle{IEEEtran}
\end{document}